\documentclass[%
 reprint,
superscriptaddress,
 twocolumn,
 sort&compress,
 amsmath,amssymb,
 aps,
 prr,
]{revtex4-2}

\usepackage{graphicx}
\usepackage{mathptmx, textcomp, float}
\usepackage{braket,amsfonts}
\usepackage[dvipsnames]{xcolor}
\usepackage[hidelinks]{hyperref}
\usepackage{upgreek}
\usepackage{eucal,enumitem}

\setlist[itemize]{itemsep=0pt}

\renewcommand{\Im}{\text{Im}}
\renewcommand{\Re}{\text{Re}}

\begin{document}

\title{Instabilities of interacting matter waves in optical lattices with Floquet driving}
\author{Andrea Di Carli}
\affiliation{Department of Physics and SUPA, University of Strathclyde, Glasgow G4 0NG, United Kingdom }
\author{Robbie Cruickshank}
\affiliation{Department of Physics and SUPA, University of Strathclyde, Glasgow G4 0NG, United Kingdom }
\author{Matthew Mitchell}
\affiliation{Department of Physics and SUPA, University of Strathclyde, Glasgow G4 0NG, United Kingdom }
\author{Arthur~La~Rooij}
\affiliation{Department of Physics and SUPA, University of Strathclyde, Glasgow G4 0NG, United Kingdom }
\author{Stefan Kuhr}
\affiliation{Department of Physics and SUPA, University of Strathclyde, Glasgow G4 0NG, United Kingdom }
\author{Charles E. Creffield}
\affiliation{Departamento de F\'{i}sica de Materiales, Universidad Complutense de Madrid,
E-28040 Madrid, Spain}
\author{Elmar Haller}
\affiliation{Department of Physics and SUPA, University of Strathclyde, Glasgow G4 0NG, United Kingdom }

\date{\today}

\begin{abstract}
We experimentally investigate the stability of a quantum gas with repulsive interactions in an optical 1D lattice subjected to periodic driving. Excitations of the gas in the lowest lattice band are analyzed across the complete stability diagram, from slow to fast driving frequencies and from weak to strong driving strengths. To interpret our results, we expand the established analysis based on parametric instabilities to include modulational instabilities. Extending the concept of modulational instabilities from static to periodically driven systems provides a convenient mapping of the stability in a static system to the cases of slow and fast driving. At intermediate driving frequencies, we observe an interesting competition between modulational and parametric instabilities. We experimentally confirm the existence of both types of instabilities in driven systems and probe their properties. Our results allow us to predict stable and unstable parameter regions for the minimization of heating in future applications of Floquet driving.
\end{abstract}
\maketitle

%***********************
% Intro
%***********************

\section{\label{sec:intro} Introduction}

Ultracold quantum gases in optical lattices have emerged as a powerful experimental platform for investigating novel quantum phenomena \cite{lewenstein2007,gross2017}. In particular the combination of optical lattices with periodically modulated driving forces provides precise control over the tunneling and band structures to create tailored lattice potentials \cite{eckardt2017}. For example, recent experiments demonstrated a dynamically driven quantum phase transition between a bosonic Mott insulator and a superfluid \cite{zenesini2009}, the generation of kinetic frustration on a triangular lattice \cite{struck2011}, the realization of artificial magnetic fields \cite{aidelsburger2011,struck2012,parker2013,aidelsburger2013, miyake2013, goldman2014}, and the creation of topological band structures \cite{jotzu2014, flaschner2016, wintersperger2020e}. Such periodically driven systems are commonly described using the Floquet formalism, which maps periodic driving to a time-independent Hamiltonian \cite{holthaus2016,eckardt2017}.

Simulating quantum many-body physics in Floquet-driven lattice potentials is challenging for interacting particles. Interactions can create instabilities and heating on short timescales, which are comparable to the modulation period and which quickly destroy the coherence of the system \cite{weitenberg2021a}. To prevent this, it is crucial to develop an understanding of excitation mechanisms and to find an optimal window for the driving frequency \cite{sun2020b}. A concept recently proposed to explain excitations in driven atomic quantum gases in lattices is parametric instability \cite{lellouch2017, lellouch2018}. Parametric instabilities occur when the driving frequency matches an excitation energy in the system \cite{bukov2015}, leading to an exponential growth of excitations and the destruction of the original quantum state. Recent experimental studies have applied this concept to analyze 1D \cite{Wintersperger2020, dupont2022} and 2D \cite{boulier2019} bosonic quantum gases.

The aim of this article is to investigate the impact of interactions on the stability of a driven quantum gas in a one-dimensional lattice potential. For a Bose-Einstein condensate of cesium atoms, we experimentally study excitation modes in the lowest lattice band and determine the system's stability across the full parameter regime of driving periods, $T_D$, and normalized driving strengths, $K$. We found it challenging to explain our measured ($T_D$,$K$)-stability diagram using only parametric instabilities. Instead, we applied the concept of modulational instabilities, which was previously studied for quantum gases with attractive interactions or negative effective mass \cite{konotop2002, modugno2004c, trombettoni2006a, cristiani2004a, fallani2004a, desarlo2005b}, and extended it to periodically driven systems. Combining both types of instabilities allowed us to develop a model that explains our measurement results and accurately predicts the system's behavior in the limits of slow and fast driving.

\begin{figure}[t]
\centering
  \includegraphics[width=0.47\textwidth]{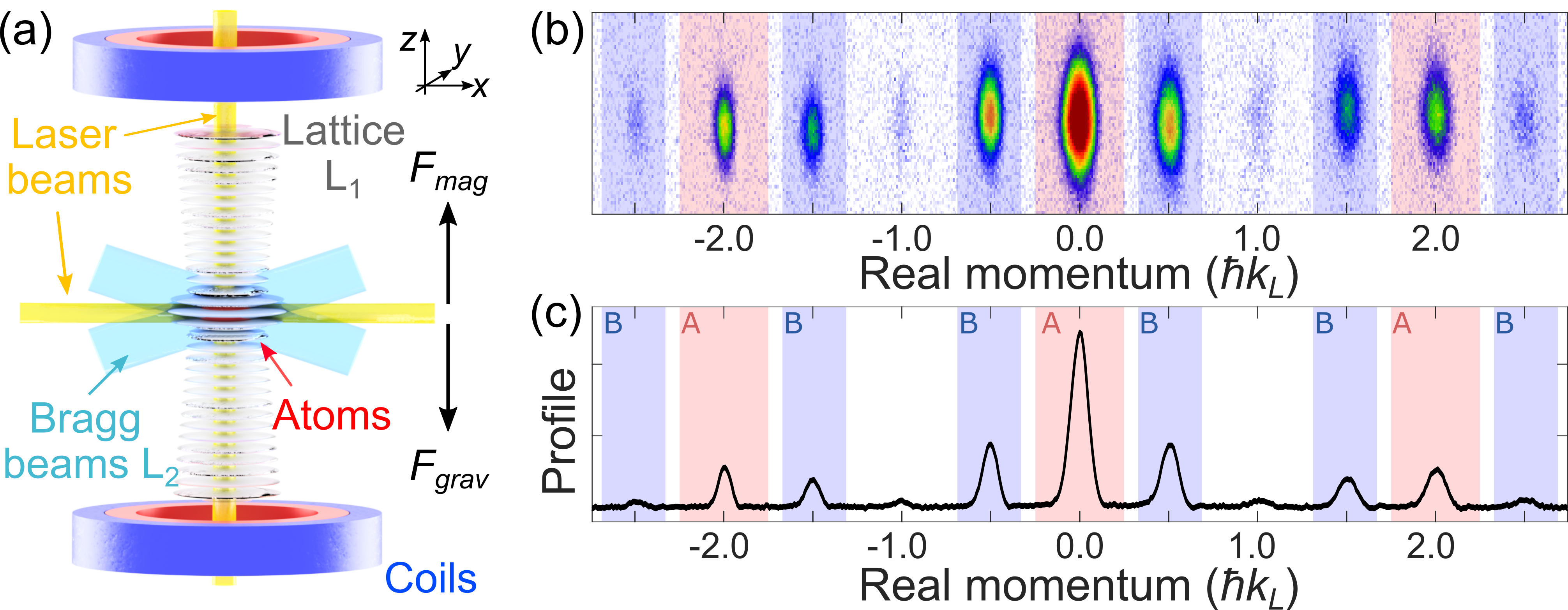} % fig1
  \vspace{-1.5ex}
  \caption{Experimental setup and absorption images. (a) Sketch of the experimental setup. (b) Absorption images showing the real momentum distribution after release from the lattice (average of 8 repetitions, images are compressed horizontally by factor 0.55). Red and blue patches indicate atoms in the carrier and in excited modes of the wave packet, respectively. (c) Integrated density profile to count the atoms in regions $A$ and $B$. \label{fig:setup}}
\end{figure}

The article is divided into three main parts. Section\,\ref{sec:Exp_MainMeasurement}
describes our experimental setup and the measurement of the system's stability diagram. For its interpretation, theoretical models of excitation modes with parametric and modulational instabilities are introduced in Sec.\,\ref{sec:InstabilitiesTheory}. We include a brief summary of excitations in static systems to provide a comprehensive understanding, before extending the description to periodic driving. Section\,\ref{sec:Exp_Measurements} contains a sequence of experiments that confirm the existence of both types of instabilities and their properties. Similar to the theory section, we begin by studying growth rates and seeding mechanisms of modulational instabilities in the static system before adding the driving force. Finally, in Sec.\,\ref{sec:InstabilityDiagram}, we combine our findings in the previous sections to explain our measurement results for the stability diagram.

%***********************
% Experiment
%***********************

\section{\label{sec:Exp_MainMeasurement} Experimental measurement of the stability diagram}

Our experimental setup is illustrated in Fig.\,\ref{fig:setup}(a). We started with a small Bose-Einstein condensate (BEC) of approximately 40,000 cesium (Cs) atoms in a vertical optical lattice, $L_1$ \cite{dicarli2019b}. The BEC was confined by an additional crossed-beam optical dipole trap with trapping frequencies $\omega_{x,y,z} = 2\pi\times (18,21,10)$\,Hz, and it was levitated by a magnetic field gradient \cite{dicarli2019a}. Lattice $L_1$ was formed by two counter-propagating laser beams with wavelength $\lambda=1064$\,nm, lattice spacing $d_L=\lambda/2$, momentum $k_L=\pi/d_L$, and depth $V=8.8\,E_r$, where $E_r$ is the recoil energy. A broad magnetic Feshbach resonance with a zero crossing at 17\,G \cite{gustavsson2008a} was used to set the s-wave scattering length, $a_s=104\,a_0$, before adiabatically loading the atoms into the lattice potential in 150\,ms.

The driving force, $F(t)$, was applied by periodically shifting the position of the sites in the lattice $L_1$. Its laser beams are independently controlled by two acousto-optical modulators with a frequency difference $\Delta \nu(t)$ between them \cite{arimondo2012a}. This frequency difference moves the lattice sites with velocity $\Delta \nu(t) d_L$ and creates an inertial force in the reference frame of the lattice. To preserve a well-defined initial momentum $k_0$ of the wave packet, we switched the force rapidly on, $F(t) = F_0 \cos(\omega_D t)$, and modulated the wave packet for a duration $t$.

For detection, we used absorption imaging to measure the momentum distribution of the wave packets after rapid switch-off of the lattice potential and after a levitated expansion time of approximately $70$\,ms (Fig.\,\ref{fig:setup}(b)). The time evolution was analysed stroboscopically at times $t$ that were multiples of $T_D=2\pi/\omega_D$. The resulting distribution of the gas in real momentum space shows the carrier wave at $k_0$ and peaks of the excitation modes at $k_0\pm q$, both with repetitions at $\pm2k_L$ (Fig.\,\ref{fig:setup}(b)). The red and blue regions indicate atoms in the main carrier wave and in excitation modes, respectively. We calculated the atom number $N_A$ in regions A that enclose the carrier wave in the momentum distribution and scaled $N_A$ to the total atom number, $N_\text{tot}$, for each image.

\begin{figure}[t]
\centering
  \includegraphics[width=0.49\textwidth]{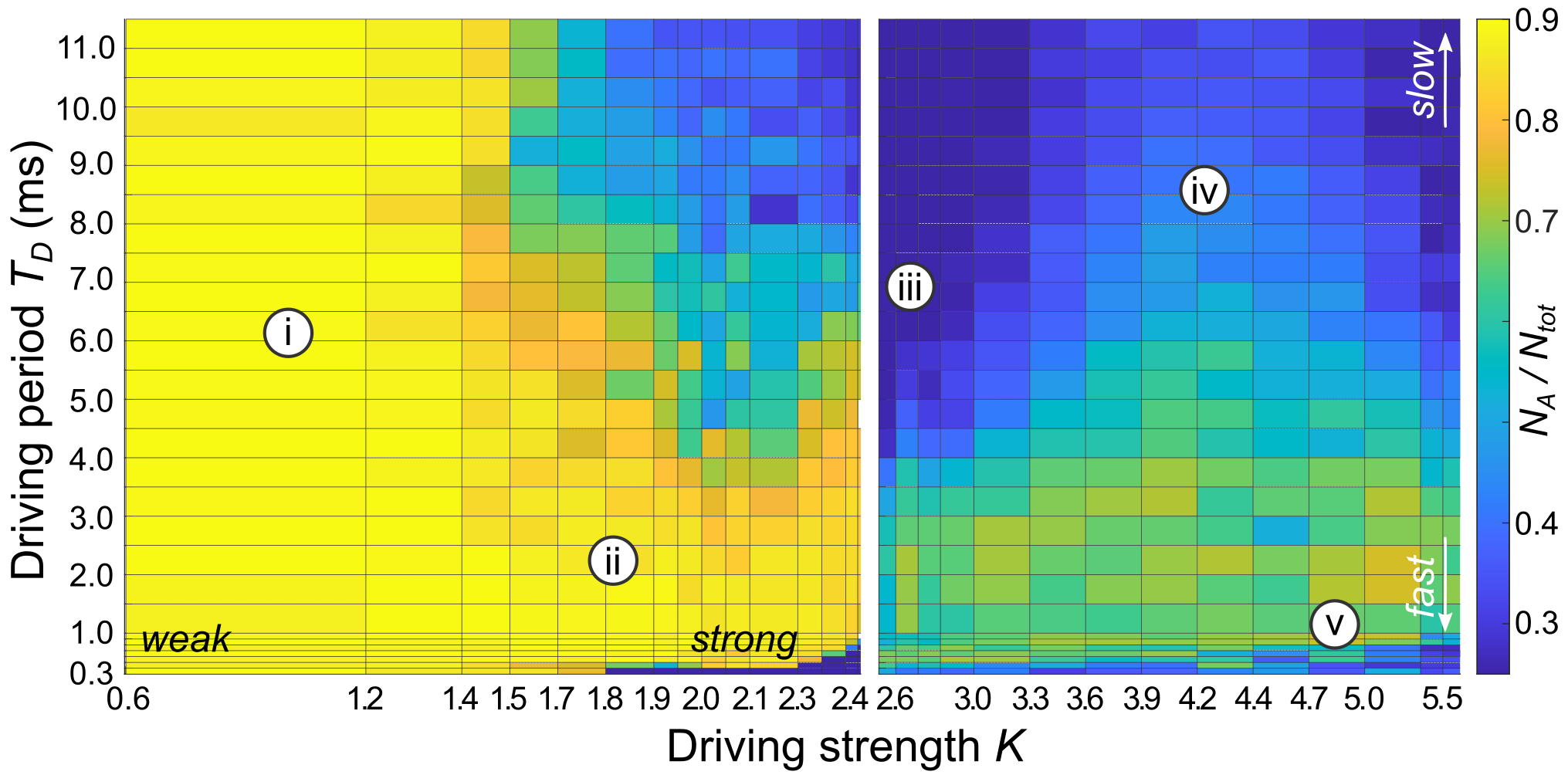} % fig1
  \caption{Measured stability diagram for a driven BEC in a 1D lattice potential. The number of atoms $N_A$ close to the initial quasimomentum of the quantum gas is measured after driving for approximately $30\,$ms with strength $K$ and driving period $T_D$. Colors from yellow to blue indicate stable to unstable regions.  \label{fig:StabilityIntro}} %System parameters $V=8.8\,E_r$, $a=105\,a_0$, $N_\text{tot}\approx 45000$.
\end{figure}

The initial wave packets for the ($T_D$,$K$)-stability diagram were prepared in the ground states of the time-averaged potentials, which are centered at $k_0 = 0$ for driving strengths $K = 0 - 2.4$ and at $k_0 = k_L$ for  $K = 2.4 - 5.5$ (see Fig.\,\ref{fig:StabilityExamples} and Supplemental Methods). The ratio of atoms in the carrier wave and the total atom number, $N_A/N_\text{tot}$, was determined after a driving duration close to 30\,ms. The exact duration was always adjusted to a multiple of $T_D$, and it was sufficiently short to reduce the effects of transversal \cite{choudhury2015, Wintersperger2020} and longitudinal excitations \cite{mitchell2021}.

The system's stability is indicated by $N_A/N_\text{tot}$ in Fig.\,\ref{fig:StabilityIntro}. Instabilities typically spread the wave packet in momentum space, and, as a result, cause a rapid reduction of $N_A/N_\text{tot}$. The measurement shows areas of high and low stability (yellow and blue colors) with clear boundaries in between. For example, the system is stable for weak driving strength $K$ and for fast driving frequencies (regions (i) and (ii) in Fig.\,\ref{fig:StabilityIntro}), but it is unstable for $K$ values close to 2.5 (region (iii)). Strong driving strengths with $K\approx4.4$ increase the system's stability again (region (iv)). For very fast driving, $T_D<0.6\,$ms (region (v)), instabilities are caused by the coupling to higher bands \cite{song2022} and will not be considered here.

The observed regions (i-iv) do not match well to predictions based on parametric instabilities \cite{bukov2015,lellouch2017}. To provide a comprehensive understanding for our measurement results, we first introduce the theoretical background for modulational and parametric instabilities in Sec.\,\ref{sec:InstabilitiesTheory} and provide experimental evidence for each mechanism in Sec.\,\ref{sec:Exp_Measurements}. We combine our findings in Sec.\,\ref{sec:InstabilityDiagram} for a detailed analysis of our measurement results in Fig.\,\ref{fig:StabilityIntro}.

%***********************
% Theory
%***********************

\section{\label{sec:InstabilitiesTheory} Theoretical models for instabilities in lattices}

We provide a brief summary of the description of excitation modes within Bogoliubov theories before introducing periodic driving into the system. An excited BEC can be described by a carrier wave with quasimomentum $\hbar k$ and a weak perturbation, $\delta \phi_k(z,t)$, \cite{modugno2004c}
\begin{align} \label{eq:perturbation}
\psi(z,t) = e^{-i\mu_k t/\hbar} e^{ikz} [\phi_{k}(z) + \delta \phi_k(z,t)].
\end{align}
Here, $\mu_k$ is the chemical potential and $\phi_{k}$ is the solution of the stationary Gross-Pitaevskii equation. We write the perturbation as a superposition of Bogoliubov excitations each with quasimomentum $\hbar q$, amplitudes $u_{kq},v_{kq}$, and energy $\hbar \omega_q(k)$
\begin{align}  \label{eq:Excitation}
\delta \phi_k(z,t) = \sum_q u_{kq}(z) e^{i(qz - \omega_q(k)t)} + v^*_{kq}(z) e^{-i(qz-\omega_q(k)t)}.
\end{align}

\begin{figure}[t]
\centering
  \includegraphics[width=0.49\textwidth]{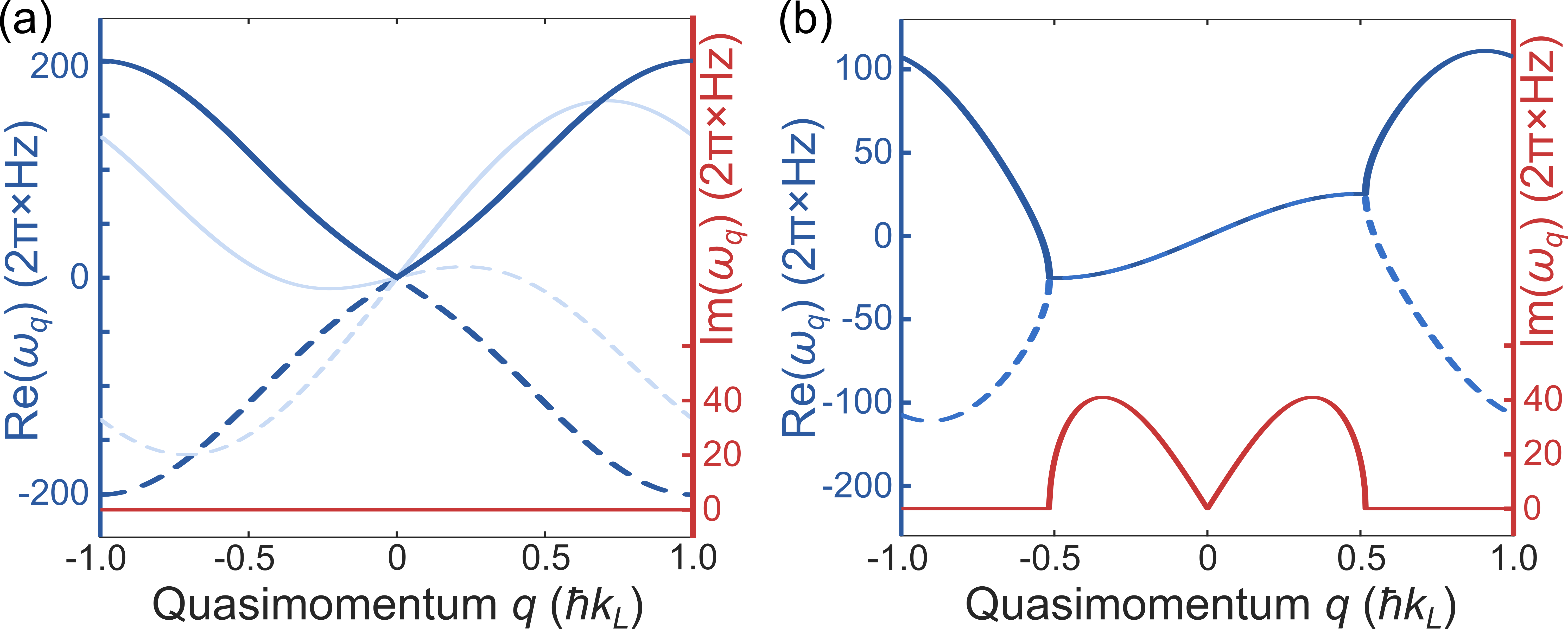}  % fig 1
  \caption{Excitation energy of phonon modes. (a) Real part of the energy for the phonon (solid lines) and anti-phonon (dashed lines) modes for $k=0$ (dark blue) and $k=0.3k_L$ (light blue) in Eq.\,(\ref{eq:ExTrombettoni}). All imaginary components are zero (red line) for parameters $V=9\,E_r$, $U/J=1$. (b) Phonon and anti-phonon modes for $k=0.9k_L$. The energy's imaginary component is non-zero when real energies of phonon and anti-phonon modes match. \label{fig:StaticEnergy}}
\end{figure}

In a lattice potential without driving, an analytical approximation for the energy of an excitation can be derived using the tight-binding ansatz and the discrete nonlinear Schr\"odinger equation \cite{wu2003a,trombettoni2006a}
\begin{multline} \label{eq:ExTrombettoni}
    \hbar \omega_{q}(k) = 2J \sin(k d_L)\sin(q d_L)  \pm 2 \\ \sqrt{ 4J^2 \cos^2(k d_L)\sin^4  \left( \frac{q d_L}{2}\right) + 2 J U  \cos(k d_L) \sin^2 \left(\frac{q d_L}{2}\right)}.
\end{multline}
Here, $J$ is the tunneling matrix element, and $U$ is an interaction coefficient that depends on the number of atoms, $N_j$, and the two-particle interaction energy at a lattice site $j$. Dipole traps in our experimental setup create a harmonic potential with a position-dependent $N_j$ along the lattice direction. For simplicity, we describe the system by its properties near the trap centre where $N_j$ is largest. The lattice potential is defined by the lattice spacing $d_L$, lattice momentum $k_L=\pi/d_L$, and a lattice depth $V$ measured in recoil energies, $E_r$. Figure \ref{fig:StaticEnergy} illustrates the energies of phonons and anti-phonons given by the plus and minus signs in Eq.\,(\ref{eq:ExTrombettoni}) \cite{modugno2004c}. For a stationary carrier medium with $k=0$, the energies are symmetric with respect to $q$ (dark blue lines in Fig.\,\ref{fig:StaticEnergy}(a)) and the system is energetically stable for $q=0$. Landau instabilities can occur when the carrier wave moves above a critical velocity (light blue lines in Fig.\,\ref{fig:StaticEnergy}(a)) and excitations, which move in the same direction, have an increased energy, while the energy is reduced for excitations with a negative $q$. The time scale for the creation of those Landau instabilities is often large and usually does not limit the stability of periodically driven BECs \cite{wu2003a}.

{\bf Modulational instabilities (MIs)} appear when the argument of the root in Eq.\,(\ref{eq:ExTrombettoni}) is negative and the energy $\hbar\omega_q$ becomes complex. As a result, excitations grow exponentially in strength with $\exp(\Gamma t)$, where the growth rate $\Gamma$ is given by $\Im(\omega_q)$. Such modulational instabilities have been studied in detail for quantum gases in lattices without periodic driving \cite{konotop2002, modugno2004c, trombettoni2006a}. They occur when the root's second term is negative and dominates the first, which happens either for attractive interactions or for a carrier wave with negative effective mass, $\cos(k d_L)<0$. As a result, MIs appear for repulsive interactions and negative effective mass when \cite{trombettoni2006a}
\begin{align} \label{eq:MIcondition}
2 J \cos^2(k d_L) \sin^2 \left( \frac{q}{2}d_L\right) <  -U \cos(k d_L).
\end{align}
For weak interactions with $U \ll 2J$, this condition is only fulfilled for small values of $|q|$ (red line in Fig.\,\ref{fig:StaticEnergy}(b)). For increasing interaction strength, this region of $q$-values with complex energy grows until it covers the entire Brillouin zone ($U=2J$) and instabilities reach maximal growth rates at $q=\pm k_L$ for $U\ge4J$.

{\bf Periodic driving} with a force $F(t) = F_0 \cos(\omega_D t)$ accelerates the wave packet to a periodic micromotion $k(t) = k_0 + (K/d_L) \sin(\omega_D t)$ through the 1st Brillouin zone. Here, $\hbar k(t)$ is the quasimomentum of the carrier wave, $K=F_0 d_L/(\hbar\omega_D)$ is the dimensionless driving strength \cite{arlinghaus2011a}, and $\omega_D=2\pi/T_D$ the driving frequency. Note that the shape of the micromotion depends only on the initial quasimomentum $\hbar k_0$ and on $K$, which is independent of $\omega_D$ when we provide the driving force by lattice shaking \cite{arimondo2012a}. For increasing values of $K$, the wave packet samples larger regions of the Brillouin zone, crossing into regions with a negative effective mass for $K=\pi/2$, until it reaches the edge of the Brillouin zone at $K=\pi$. For even larger values of $K$, Bragg-scattering occurs at the band edge and folds the micromotion back into the 1st Brillouin zone. To simplify the description, we refer in the article to weak and strong driving strengths for $K$ below and above $\pi/2$.

The time evolution of an excitation mode can be described by the Bogoliubov-de Gennes equations for periodic driving \cite{creffield2009}. Without interactions, the components $u$ and $v$ evolve with energy $\epsilon_\pm(q,t)$, while the interaction energy $U$ couples the evolution with
\begin{align} \label{eq:BogoliubovGennes}
 i\partial_t \begin{pmatrix} u_q \\ v_q \end{pmatrix} = \begin{pmatrix} \epsilon_+(q,t) + U & U \\ -U & - \epsilon_-(q,t)-U \end{pmatrix}  \begin{pmatrix} u_q \\ v_q \end{pmatrix}
\end{align}
where
\begin{align}
     \epsilon_\pm(q,t) = 4 J \sin\left(\frac{q}{2}d_L\right) \sin\left(\frac{q}{2}d_L \pm k_0d_L \mp K \sin(\omega_D t) \right). \nonumber
\end{align}
The index $k$ is omitted for $(u_{q},v_{q})$ by assuming that the momentum of the carrier wave is $k$ without driving and that we probe the system at multiples of the driving period with $k=k_0$ for $K>0$.

\begin{figure}[t]
\centering
  \includegraphics[width=0.49\textwidth]{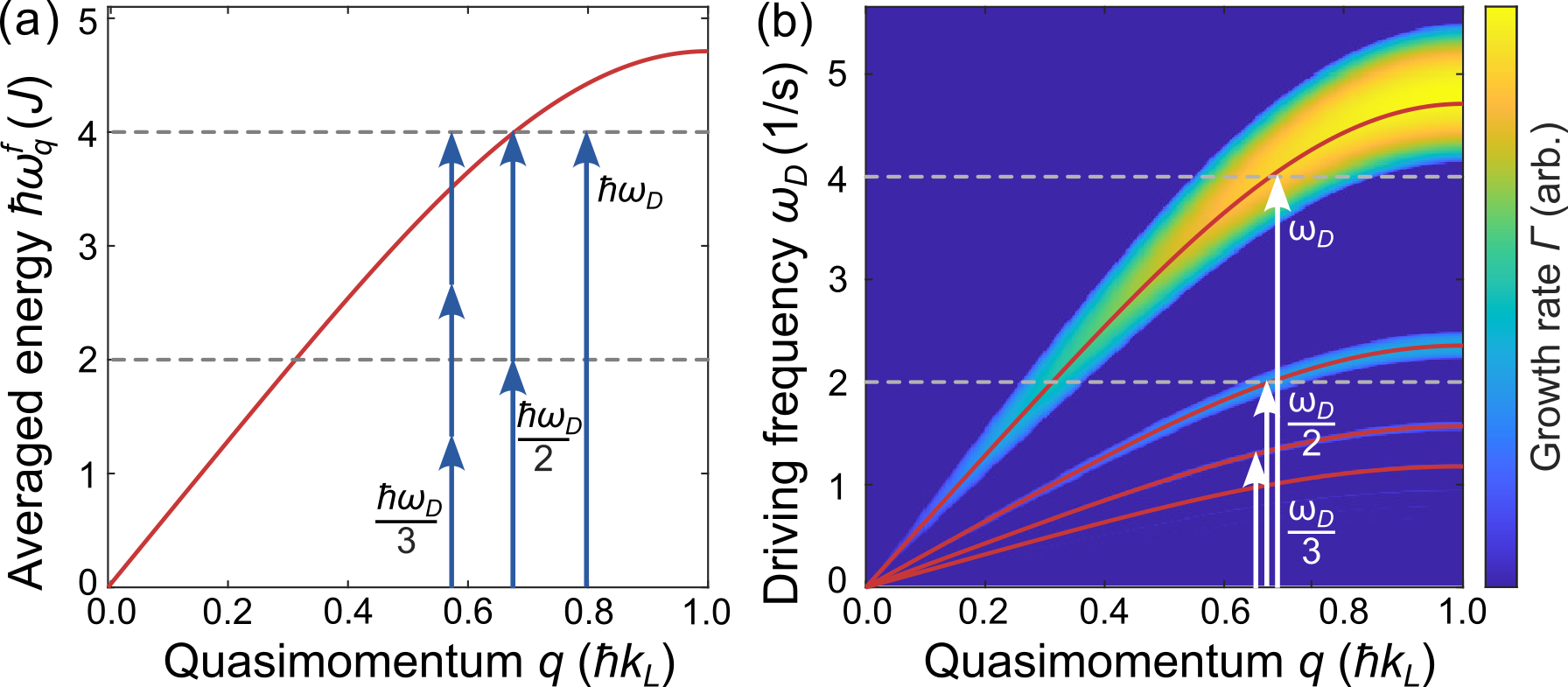}  % fig 2
  \caption{Parametric resonances. (a) Calculated excitation energy $\hbar \omega^f_q$ of modes with momentum $q$ ($V=8\,E_r$, $U=4J$, $K=1.4$). Resonances occur when integer multiples of the driving frequency $\omega_D$ match $\omega^f_q$. (b) Calculated growth rate $\Gamma$ for excitations modes with momentum $q$ and driving frequency $\omega_D$. Red lines indicate the frequency matching condition $\omega^f_q =n\omega_D$. \label{fig:IllustratePIs}}
\end{figure}

The Bogoliubov-de Gennes equations can be simplified for slow and fast driving frequencies. As a reference, we use the largest energy that an excitation mode acquires due to the micromotion of the carrier wave, $\hbar \omega_\text{max} = 2 \sqrt{ 4J^2 + 2 J U }$. This energy provides an upper limit for the oscillation frequencies of the $u,v$ components during one driving period. For faster frequencies, $\omega_D\gg\omega_\text{max}$, the response of the system is too slow to follow the drive, and $\epsilon_\pm(q,t)$ can be replaced by time-averaged values \cite{creffield2009}. Diagonalizing the resulting matrix in Eq.\,(\ref{eq:BogoliubovGennes}) provides the energy of phonon and anti-phonon modes in the fast driving limit
\begin{multline}
\hbar \omega^f_q(k_0,K) = 2 J_\text{eff} \sin(k_0 d_L) \sin(q d_L) \pm \sqrt{4J_\text{eff} \sin^2 \left( \frac{q d_L}{2}\right)} \\
\overline{ \left( 4J_\text{eff} \cos^2(k_0 d_L)\sin^2\left(\frac{q d_L}{2}\right) +2U\cos(k_0 d_L) \right) }  \label{eq:FastDrivingLimit}
\end{multline}
where $J_\text{eff}=J J_0(K)$ is the renormalized tunneling matrix element and $J_0$ is the zeroth-order Bessel function.

Equation\,(\ref{eq:FastDrivingLimit}) is similar to the effective Bogoliubov dispersion relation in \cite{creffield2009, bukov2015, lellouch2017}, but extended by the parameter $k_0$. This extension allows us to use the same representation for the effective dispersion relation in a driven system as in a non-driven system, i.e., Eq.\,(\ref{eq:ExTrombettoni}) maps directly to Eq.\,(\ref{eq:FastDrivingLimit}) for fast driving and a stroboscopic description, by replacing $J$ with   $J_\text{eff}$, and $k$ with the initial momentum $k_0$. This result implies that the concept of MIs and their properties can be well extended to systems with fast driving. As in the static case, MIs appear in the time-averaged system when $\omega^f_q(k_0,K)$ has complex components, i.e., the system is unstable for
\begin{align} \label{eq:MIconditionDriven}
2 J_\text{eff}^2 \cos^2(k_0 d_L) \sin^2 \left( \frac{q d_L}{2}\right) < - J_\text{eff} U \cos(k_0 d_L),
\end{align}
which can be fulfilled if one of the parameters $J_\text{eff}, \cos(k_0 d_L),$ or $U$, is negative. For example, the initial evolution of a BEC after quenching the sign of $J_\text{eff}$ \cite{lignier2007} can be interpreted as the growth of modulational instabilities.

\begin{figure}[t]
\centering
  \includegraphics[width=0.49\textwidth]{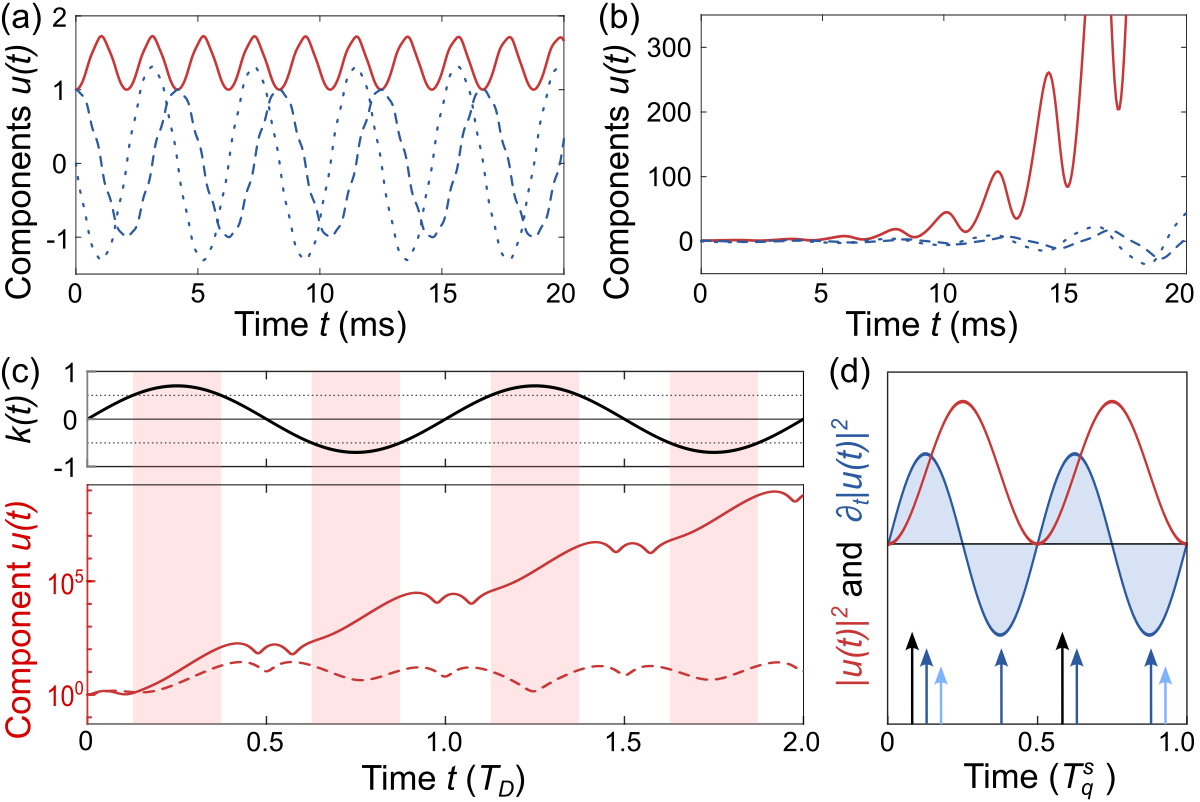}  % fig 3
  \caption{Calculated time evolution of an excitation mode $(u=1,v=0)$. (a) Components of $u(t)$ oscillate with the mode's intrinsic frequency $\omega^f_q$ in the fast driving limit. Dotted, dashed, and red lines show $\Im[u(t)]$, $\Re[u(t)]$, $|u(t)|^2$, for parameters $\omega_D=4\omega^f_q$, $K=1.4$, $q=\hbar k_L$, $U=10J$. (b) A parametric instability appears when $\omega_D$ matches this intrinsic frequency and the excitation mode grows exponentially. (c) Modulational instabilities require the micromotion (black line top panel) to cross into regions with negative effective mass (red patches). The system is modulationally unstable for $T_D=11.3\,$ms (solid red line), but stable for $T_D=8.3$\,ms (dashed red line) with parameters ($K=2.2$, $k_0=0$, $U=10J$). (d) Sketch of the oscillation of $|u|^2$ (red line) and its derivative (blue line). The system is stable when the effective mass turns negative (arrows) in intervals with alternating slope of $|u|^2$. \label{fig:TimeEvolution}}
\end{figure}

Slow driving with $\omega_D\ll\omega_\text{max}$ requires the opposite approach. The oscillations of the $u,v$-components and the system's response are fast compared to $T_D$ and the excitation energies are given by Eq.\,(\ref{eq:ExTrombettoni}) for each moment in time. Large growth rates with $\Gamma\gg\omega_D$ limit the application of the model as excitation modes can grow quickly and break the perturbative ansatz in Eq.\,(\ref{eq:BogoliubovGennes}) before even a single driving cycle is completed. However, we find it helpful to define the average energy of an existing excitation as the system cycles through one micromotion
\begin{align} \label{eq:SlowDrivingLimit}
  \hbar \omega^s_q(k_0,K)=  \frac{1}{T_D} \int_0^{T_D} \hbar \omega_{q}(k(t)) dt,
\end{align}
where $k(t)$ depends on $k_0$ and $K$. The values of $\omega^s_q$ and $\omega^f_q$ are closely matched for weak driving strengths but deviate when the micromotion crosses into regions with negative effective mass and $\omega^s_q(k_0,K)$ acquires imaginary components. Both parameters, $\omega^s_q$ and $\omega^f_q$, have limitations when describing the interesting regime of intermediate driving frequencies with comparable values for $\omega_D$ and $\omega_\text{max}$.

{\bf Parametric instabilities (PIs)} occur in this intermediate frequency regime when $\omega_D$ resonantly matches excitation energies in the system \cite{bukov2015, lellouch2017}. We illustrate this condition in Fig.\,\ref{fig:IllustratePIs}(a) by comparing $\omega_D$ to $\omega^f_q$. Parametric resonances occur at quasimomenta for which $\omega^f_q$ is equal to integer multiples of $\omega_D$ (blue arrows in Fig.\,\ref{fig:IllustratePIs}(a)). This concept works well to explain the numerically calculated growth rate for weak driving. We propagate Eq.\,(\ref{eq:BogoliubovGennes}) for $u$ and $v$ over one period, diagonalize the resulting matrix, and use its eigenvalues to determine $\Gamma_q(K,T_D)$ \cite{creffield2009}. The regions with large values of $\Gamma_q$ (green-to-yellow colors in Fig.\,\ref{fig:IllustratePIs}(b)) match well to the red lines that indicate the resonance condition $\omega^f_q=n\omega_D$ with integer $n$. Similar fractional excitation spectra occur in a tilted lattice \cite{haller2010} and for interband excitations \cite{reitter2017b}.

\begin{figure}[t]
\centering
  \includegraphics[width=0.49\textwidth]{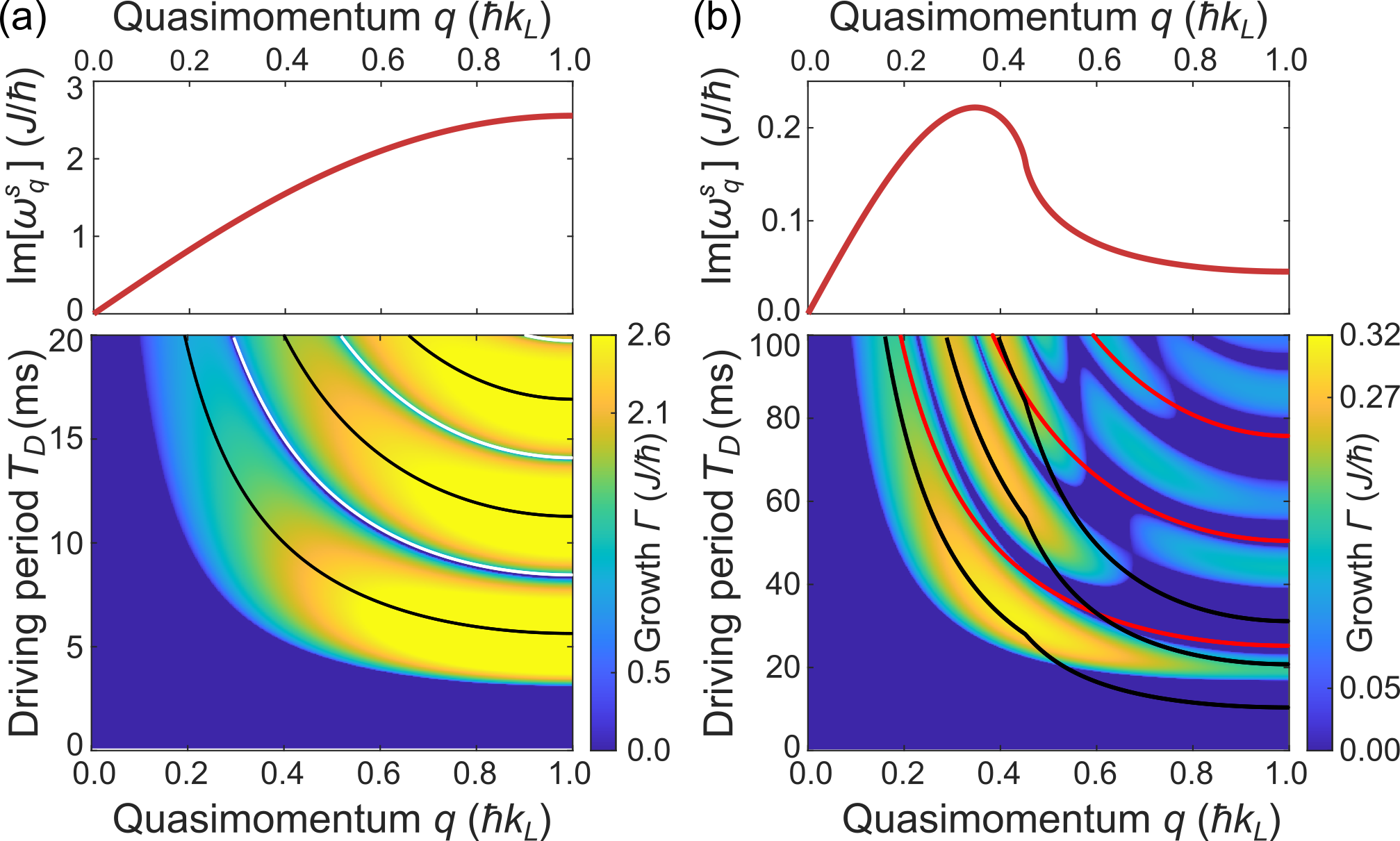} % fig 4
  \caption{Calculated growth rate for strong driving strength. (a) Imaginary component of time-averaged phonon energy $\hbar \omega_q^s$ (top) and the growth rate (bottom) for strong interactions $U=10J$ ($K=2.2$, $V=8\,E_r$). Black and white lines indicate integer and half-integer multiples of $T^s_q$.  (b) Same panels for weak interactions with $U=0.5J$. The red lines and black lines indicate multiples of $2\pi/\omega^f_q$ and $T^s_q$. \label{fig:ResonancesStrongDriving}}
\end{figure}

We further investigate the cause of the instabilities by studying the time evolution of the excitation mode ($u=1,v=0$) over a few oscillation periods using Eq.\,(\ref{eq:BogoliubovGennes}). In the case of fast driving and weak driving strength, the real and imaginary components of $u(t)$ oscillate with frequency $\omega^f_q$ (blue lines in Fig.\ref{fig:TimeEvolution}(a)) and $|u|^2$ oscillates with frequency $2\omega^f_q$ (red line in Fig.\ref{fig:TimeEvolution}(a)). Parametric instabilities arise when the driving frequency approaches $\omega^f_q$, causing $|u|^2$ to diverge exponentially (Fig.\ref{fig:TimeEvolution}(b)). In contrast, MIs only occur for strong driving strength with $K>\pi/2$ when the micromotion crosses into regions of the Brillouin zone with negative effective mass. For instance, Fig.\,\ref{fig:TimeEvolution}(c) illustrates the micromotion and critical regions with negative effective mass with a black line and red patches, respectively. The excitation mode grows whenever the micromotion crosses through such critical regions (solid red line). It would seem reasonable to expect that interacting systems are always unstable for strong driving. However, we also observe narrow intervals of $\omega_D$ with large yet stable oscillations (dashed red line).

To understand the cause of those narrow stable intervals, we analyze the growth of excitations for strong interactions and $K$ close to 2.4 (Fig.\,\ref{fig:ResonancesStrongDriving}(a)). The parameter regime is chosen because the time-averaged excitation energy $\hbar \omega^s_q$ provides a good approximation for the oscillation of $u(t)$ with period $T^s_q = \Re[2\pi/\omega^s_q]$ and growth $\Im[\omega^s_q]$. We observe a large growth rate for all values of $q$ (top panel), suggesting that the system is always unstable due to MIs. However, our numerical calculations using Eq.\,(\ref{eq:BogoliubovGennes}) reveal the presence of stable regions (blue colors in bottom panel). Stability is achieved for $T_D=nT^s_q/2$ (white lines), where $n$ is an odd integer, and for $T_D<T^s_q/2$ in the fast driving limit.

Those stable intervals in the driving frequency are caused by an interplay between the periodic growth of MIs and the oscillation of $|u|^2$ with period $T^s_q/2$. Crossings into critical regions with negative effective mass cause the system to be periodically unstable. Surprisingly, this can reduce the oscillation amplitude of the mode, when the crossing occurs on a downward slope of $|u|^2$. The stability of the system depends on the number of crossings into critical regions of the Brillouin zone per interval $T^s_q$. It is unstable for two crossings (black arrows in Fig.\,\ref{fig:TimeEvolution}(d)),  but stable for at least four crossings (dark blue arrows) when the growth during upward and downward slopes of $|u|^2$ cancels out. This is the origin of the fast driving limit. The stability at the white lines occurs when the micromotion crosses into critical regions at alternating slopes of the $|u|^2$ oscillation, such as at $T_D = 3T^s_q/2$ (light blue arrows). A similar analysis is challenging for weak interactions with $U\approx J$ as neither $\omega^s_q$ nor $\omega^f_q$ are well suited to describe relevant time scales in the system (black and red lines in Fig.\,\ref{fig:ResonancesStrongDriving}(b)). The growth of MIs is stronger for small $q$-values than for large (top panel), which results in a complicated competition between MIs and PIs (bottom panel). MIs control the growth of excitations in the center of the Brillouin zone, whereas PIs dominate for large $q$-values, with a complicated pattern of stable and unstable regions in between.

{\bf Comparing PIs and MIs.} Modulational instabilities are well suited to describe the behavior of excitation modes in the fast and slow driving limits. They are caused by intrinsic properties of the medium, e.g., attractive interactions or a negative effective mass in the Brillouin zone, which already make the non-driven system unstable. Driving and the resulting micromotion just switch instabilities periodically on and off. For the fast driving limit, time averaging over the micromotion provides an intuitive description of phonon energies, stability criteria, and growth rates that is formally identical to the non-driven system.

Parametric instabilities arise when the driving frequency resonantly couples with an excitation mode, causing a stable system to become unstable. As a result, PIs can exist only for intermediate driving frequencies with comparable excitation energies. This regime of intermediate driving frequencies is most interesting because PIs and MIs can exist simultaneously. Parametric instabilities make a mostly stable system unstable for resonant driving frequencies (Fig.\,\ref{fig:IllustratePIs}(b)), while systems with MIs are mostly unstable but show narrow frequency intervals with stable modes (Fig.\,\ref{fig:ResonancesStrongDriving}(a)).

%***********************
% Experiment
%***********************

\section{\label{sec:Exp_Measurements} Experimental characterization of MIs and PIs}

This section provides a sequence of experimental measurements that demonstrate the existence of both types of instabilities and study their properties. We found it helpful to probe the properties of excitation modes separately for non-driven and driven systems, and for MIs and PIs, before combining our results for the analysis of the main stability measurement in Sec.\,\ref{sec:InstabilityDiagram}. For non-driven systems, we demonstrated the growth of excitation modes when the carrier wave has a negative effective mass (Sec.\,\ref{sec:Exp_GrowthMI}), and we observed the decay of modes in stable regions of the Brillouin zone (Sec.\,\ref{sec:Exp_GrowthStabMI}). Combining both concepts allowed us to analyse the observed evolution of excitation modes in driven systems when the micromotion cycles through the Brillouin zone (Sec.\,\ref{sec:Exp_MIDriven}). Demonstrating the existence of PIs was challenging, as we never observed the formation and growth of excitations in parameter regimes which allow for PIs but not for MIs. Instead, we created excitation modes directly with the help of Bragg scattering pulses and measured the mode's reduced decay due to PIs (Sec.\,\ref{sec:Exp_PIDriven}).

\begin{figure}[t]
\centering
  \includegraphics[width=0.49\textwidth]{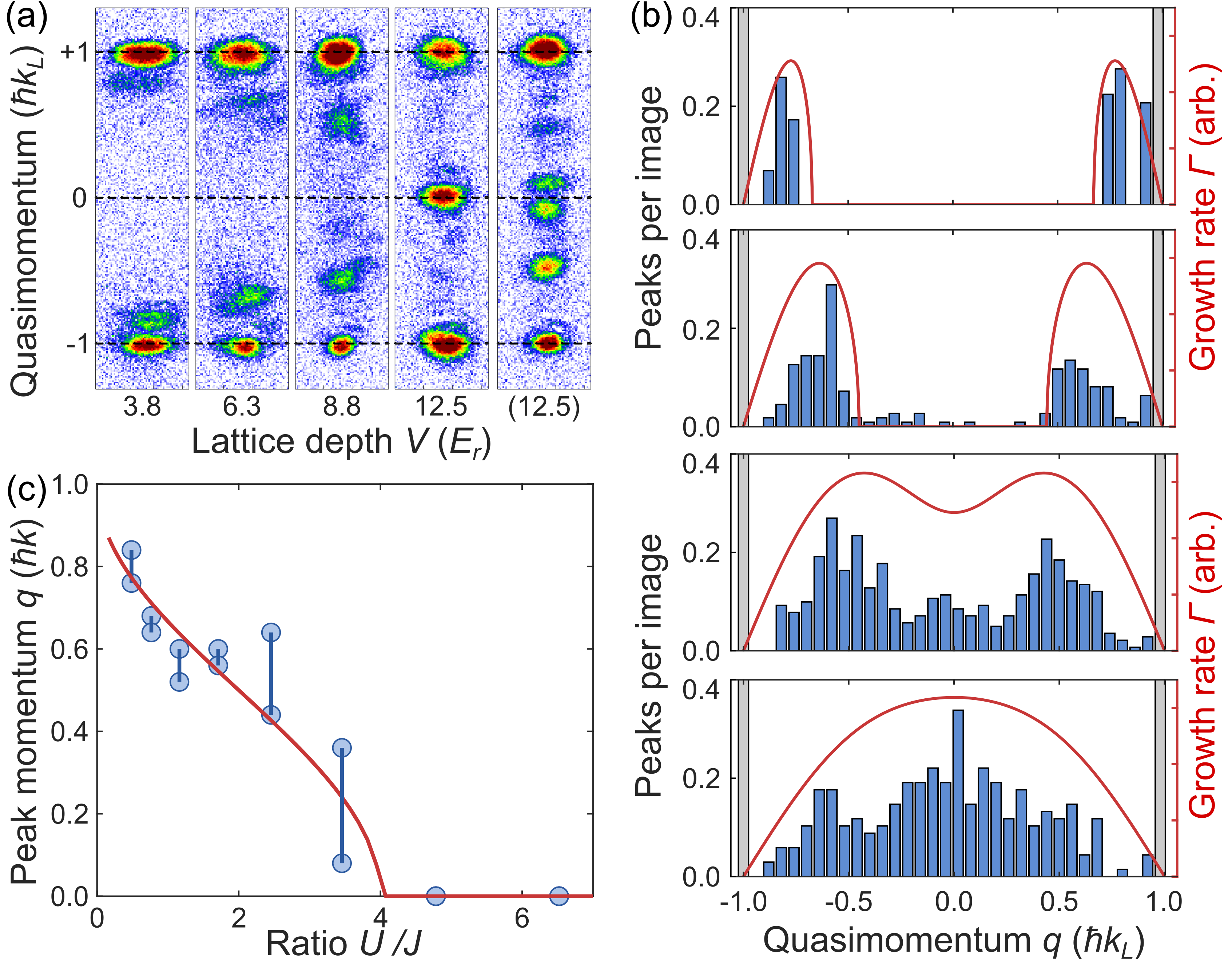} % fig7
  \caption{Momentum distribution of MIs without driving. (a) Example images of MIs for different lattice depths $V$ and $k=k_L$ after approximately $10$\,ms ($a=6\,a_0$, $N=3\times10^4$). We observe images with matching pairs of excitation modes ($V=3.8-12.5\,E_r$), but also images with multiple modes ($V=12.5\,E_r$ right). (b) Histogram of peak positions for increasing lattice depths and $U/J=0.5, 1.2, 2.5, 6.5$ (top-to-bottom). Red lines indicate $\Gamma=\Im(\omega_q)$. (c) The measured peak positions of the histograms match well to the predicted positions in Eq.\,(\ref{eq:MIPosition}) (red line). \label{fig:StaticMIPosition}} %(Function plotGrowthFunction.m) % data 19.05.22
\end{figure}

\subsection{\label{sec:Exp_MIstatic} Modulational instabilities without periodic driving}

Without periodic driving, modulational instabilities were experimentally studied in detail during the 2000s \cite{cristiani2004a, fallani2004a, desarlo2005b, campbell2006b}. For example, MIs can grow in homogeneous superfluids with attractive interactions, leading to the formation of soliton trains \cite{nguyen2017}, and they can occur for repulsive interactions in lattice potentials with quasimomenta that have a negative effective mass \cite{cristiani2004a, fallani2004a}. In this section, we experimentally measure properties of MIs with relevance to the driven system, such as the momentum distribution, growth rate, and time evolution.

\subsubsection{\label{sec:Exp_MomentumMI} Momentum distribution.}

We first studied the momentum distribution of excitation modes that develop due to MIs for carrier momentum $\hbar k_L$. We accelerated the wave packet with a magnetic field gradient to the carrier momentum, waited for approximately 10\,ms and measured its quasimomentum distribution. The resulting momentum profiles (Fig.\,\ref{fig:StaticMIPosition}(a)) show the carrier wave at $k_L$ and two peaks, which are caused by pairs of excitation modes with opposite quasimomenta $(+\hbar q,-\hbar q)$ \cite{vogels2002}.

The momentum of the most unstable mode, $q_\text{mum}$, can be calculated using Eq.\,(\ref{eq:ExTrombettoni}),
\begin{align} \label{eq:MIPosition}
     \cos\left(\frac{q_\text{mum}d_L}{2}\right) = \left\{
    \begin{array}{ l l }
    -\sqrt{1 - \frac{U}{4 J |\cos(k d_L)|}} ,  & \textrm{if\,} U \leq 4 J|\cos(k d_L)| \\
    0,                                                     & \textrm{if\,} U > 4J|\cos(k d_L)|
  \end{array}
  \right.
\end{align}
Depending on the ratio $U/J$, the value of $q_\text{mum}$ shifts from the centre of the Brillouin zone ($U\ll J$) to the edge of the BZ when $U$ is equal to the band width $4J$. Note that the quasimomentum $\hbar q$ of an excitation mode is defined in the reference frame of the carrier wave. As a result, excitation modes have a momentum $\hbar k\pm \hbar q$ in the lattice frame, and they appear in the images at the edge of the Brillouin zone for $q=0$ and at the centre for $q=k_L$.

In addition to the expected peaks at momentum $\pm q_\text{mum}$, we observed images with more than two modes and shot-to-shot fluctuations of their momenta (see $2^\text{nd}$ example image for $V=12.5\,E_r$ in Fig.\,\ref{fig:StaticMIPosition}(a)). To account for these fluctuations, we used a statistical analysis of the MIs. We took 100-150 images, applied a peak-finding algorithm to detect excitation modes and calculated histograms of their positions, Fig.\,\ref{fig:StaticMIPosition}(b). The shapes of the histograms match well with the predicted growth rates $\Gamma_q=\Im(\omega_q)$ for increasing interaction strengths $U/J$ (solid lines in Fig.\,\ref{fig:StaticMIPosition}(b)). The absolute values of the growth rates will be discussed in Sec.\,\ref{sec:Exp_GrowthMI}.

For the measurement of $q_\text{mum}$, we determined the two peaks with the most likely quasimomentum for positive and negative values of $q$ (Fig.\,\ref{fig:StaticMIPosition}(c)). Connected pairs of data points indicate the momenta of the two peaks per histogram. Histograms with $U/J>4$ have a single peak at $q=0$. Our data matches well to the predicted values for $q_\text{mum}$ in Eq.\,(\ref{eq:MIPosition}), and we expect that the observed $U/J$-dependence of $q_\text{mum}$ can be extended to driven systems. However, there is a discrepancy with the models used in previous studies \cite{lellouch2017, boulier2019, Wintersperger2020}. In our measurements, the most unstable mode does not dominate the system's time evolution, but it is merely the most likely mode in a distribution. While, in a scenario of infinite carrier medium and infinite observation time, the most unstable mode would outgrow all others, such conditions do not exist in our system.

\subsubsection{\label{sec:Exp_GrowthMI} Growth rates.}

\begin{figure}[t]
\centering % first attempt 22.05
  \includegraphics[width=0.49\textwidth]{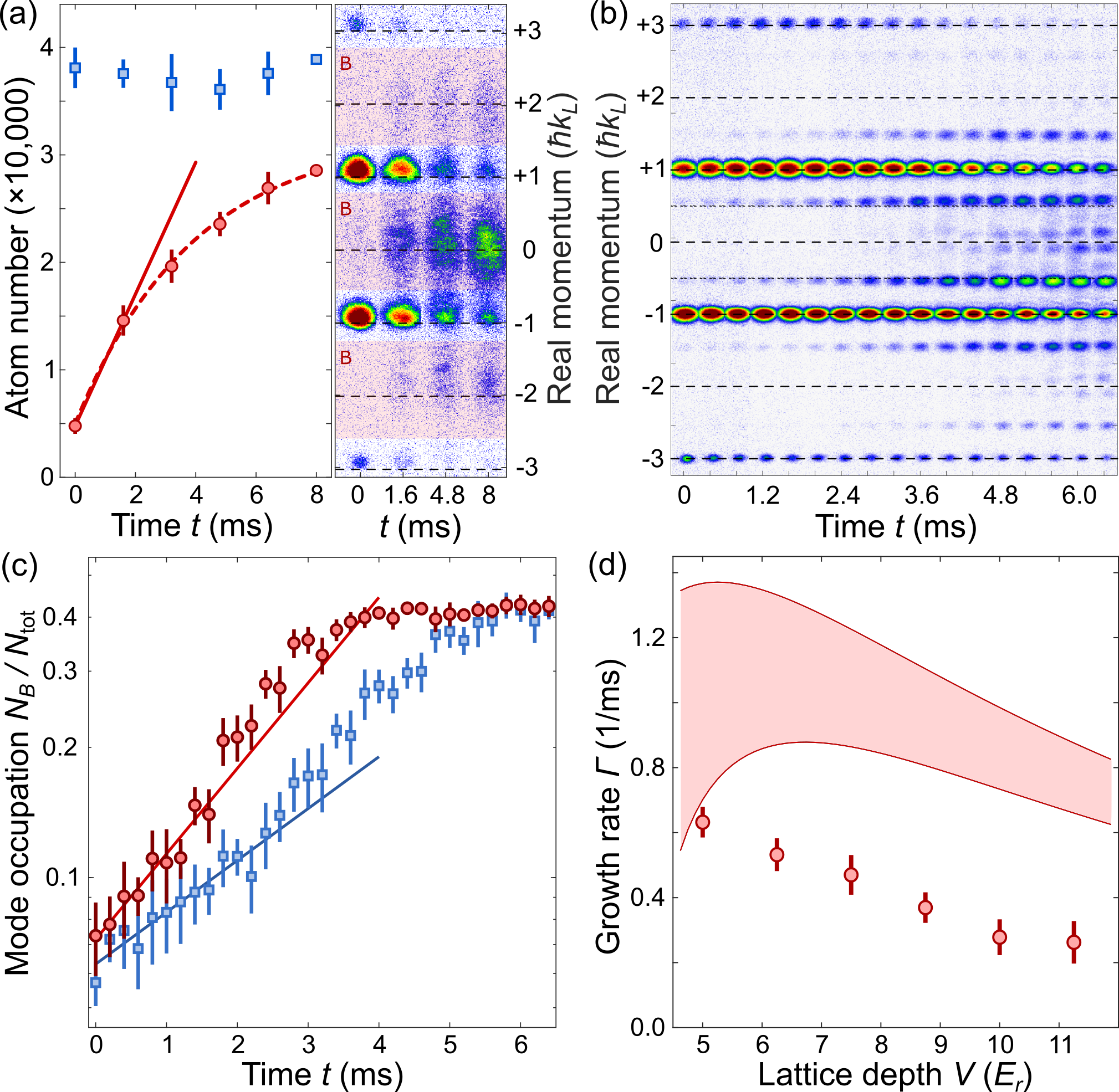} % fig8
  \caption{Growth rate of MIs without driving. (a) Time evolution of excitations with $N_B$ atoms at the centre of the BZ, (red circles, red patches) and total atom number, $N_\text{tot}$ (blue squares), $k=0.95 k_L$, $V=10\,E_r$, $a=90\,a_0$. Lines are added to guide the eye. (b) Absorption images of the time evolution after seeding excitations at $q=\pm0.46$ ($k=k_L$, $V=10\,E_r$, $a=14\,a_0$, $N_\text{tot}\approx 41,000$, $U/J = 8.9$, 5 repetitions). (c) Relative atom number in seeded excitation modes for $V=10\,E_r$ (blue squares) and $V=6.3\,E_r$ (red circles). The lines are exponential fits to the data points with $t<1.5\,$ms. (d) Measured growth rate $\Gamma_q$ (red circles) and range of expected rates $\Im(\omega_q)$ assuming lattice site occupations between 60\% and 100\% of the peak value (red patch).
  \label{fig:StaticMIGrowth}} % data (a) 22.05.22 curve a-b\exp(-c t), (b) 23.07.22 peak 2400 atoms, nu/J = 8.9, (d) 28/07
\end{figure}

The growth rate $\Gamma_q$ of excitations is difficult to measure directly due to their random occurrence. To illustrate this challenge, we used a magnetic field gradient that accelerated the wave packet to momentum $\hbar k$, waited for a variable hold time, and evaluated the momentum profile using absorption imaging. We observed that excitation modes grow predominantly near the center of the Brillouin zone (region B in Fig.\,\ref{fig:StaticMIGrowth}(a)), resulting in an increase in atom number $N_B$ in that region. Instead of exhibiting exponential growth as expected, the increase in $N_B$ shows a convex curvature. This curvature results from the simultaneous or sequential growth of multiple excitation modes. Similar non-exponential growth patterns were also reported for a driven system in \cite{boulier2019}, and for non-driven systems when using atom loss as an indicator for mode growth~\cite{fallani2004a}.

To enhance the predictability of MIs, we seeded the modes directly using Bragg pulses generated by lattice $L_2$ (Fig.\,\ref{fig:setup}(a)). The pulse intensity and duration were adjusted to create a seed with 5-8\% of the atoms in the carrier wave. The absorption images of the time evolution show an initial increase in the excited modes for 4\,ms, followed by coupling to higher order modes at $\pm2q$ (Fig.\,\ref{fig:StaticMIGrowth}(b)). To evaluate the growth, we measured the relative atom numbers $N_B/N_\text{tot}$ in the excitation modes (Fig.\,\ref{fig:StaticMIGrowth}(c)). Initially, the seeded modes experience exponential growth that is followed by saturation or even a decrease, which is due to the limited number of atoms and the growth of higher-order excitations. When determining the exponential growth rate within the first 1.5\,ms, the measured values of $\Gamma_q$ are significantly smaller than $\Im(\omega_q)$, which are the values suggested by Eq.\,(\ref{eq:ExTrombettoni}) (Fig.\,\ref{fig:StaticMIGrowth}(d)). This difference between the measured and expected growth rates increases further when we consider that the measurement probes the density of the excitation modes, which should show twice the growth rate \cite{boulier2019}.

We speculate that this discrepancy between predicted and measured growth rates is caused by different initial states of the excitation modes. $\Im(\omega_q)$ is calculated for the eigenstates $(u_q,v_q)$ of the matrix in Eq.\,(\ref{eq:ExTrombettoni}). However, these states may not match the modes created by Bragg scattering or by random seeding, leading to initial oscillations (red line Fig.\,\ref{fig:TimeEvolution}(a)) and an inaccurate measurement of exponential growth. In addition, the calculated growth rates also become unreliable for longer hold times, as the perturbative ansatz in Eq.\,(\ref{eq:perturbation}) is no longer applicable for large excitation modes.

\subsubsection{\label{sec:Exp_GrowthStabMI} Decay rates.}

\begin{figure}[t]
\centering
  \includegraphics[width=0.49\textwidth]{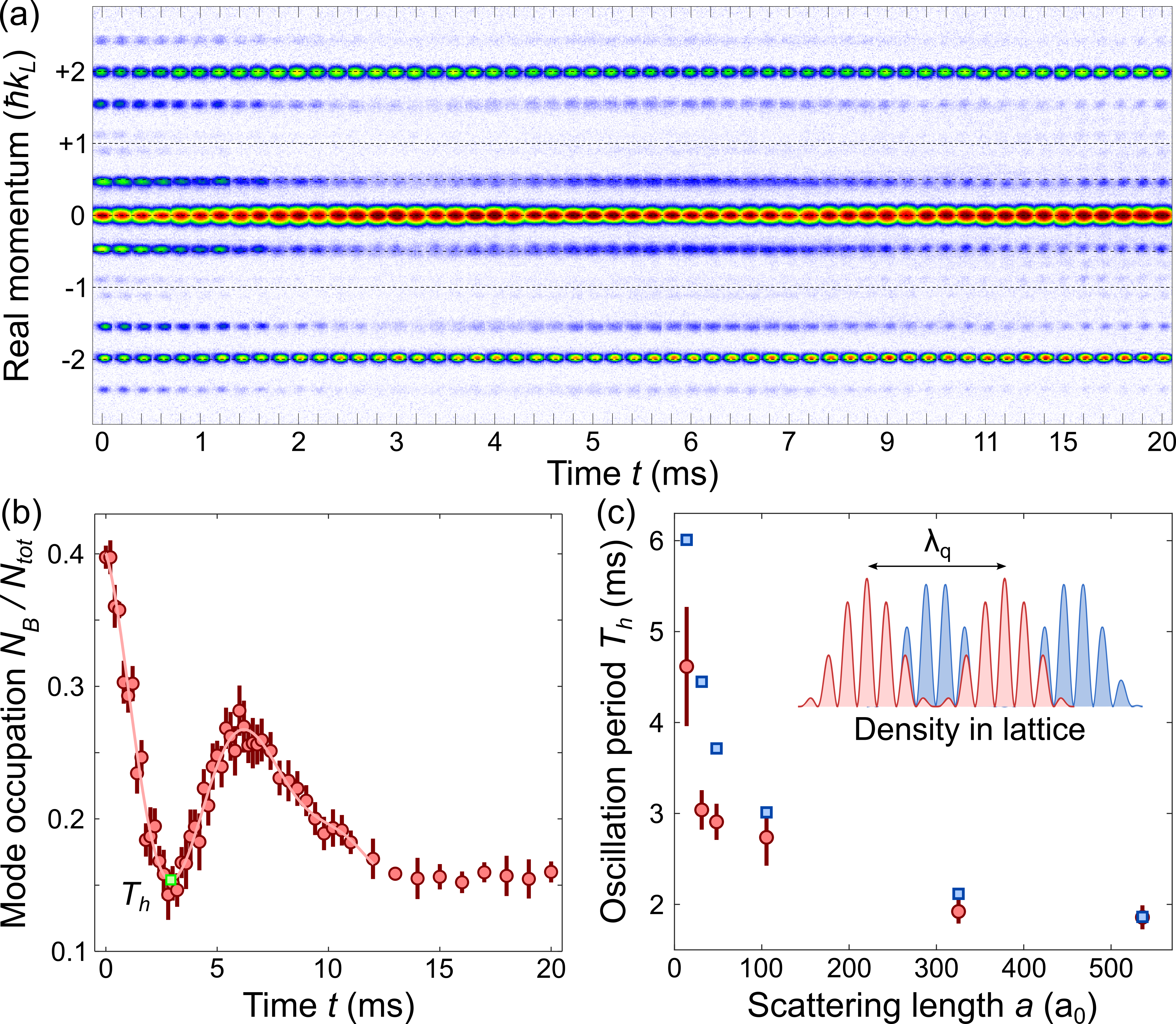} % fig 8
  \caption{Decay of excitations in stable regions without driving. (a) Absorption images of the time evolution of excitation modes $k=0$ ($a_s=31\,a_0$, $V=10\,E_r$, $q=0.47\,k_L$). The excitation modes are created by a Bragg pulse which excites approximately $40\,\%$ of the atoms. Images are centered to the $q=0$ peak to averaged 8 repetitions. (b) Fraction of atoms $N_B/N_\text{tot}$ in excited mode for data in (a). We determine the time at the minimum $T_h$ by locally fitting a 2nd order polynomial. (c) $T_h$ for various scattering lengths (red circles) and period $2\pi/\omega_q$ in Eq.\,(\ref{eq:ExTrombettoni}) (blue squares). The inset shows a sketch of two density profiles (red and blue color) for a pair of excitation modes with $\pm q$. \label{fig:StaticMIHealing}} % data 25.07.22 and 26.07
\end{figure}

Before discussing driven systems in the next section, it is instructive to study the evolution of existing excitation modes in a stable region of the Brillouin zone. Compared to Sec.\,\ref{sec:Exp_GrowthMI}, we used stronger Bragg pulses that excited approximately $40\%$ of the carrier wave to momentum $\pm q$ and prepared the carrier wave at $k=0$. The resulting time evolution (Fig.\,\ref{fig:StaticMIHealing}(a,b)) shows a fast oscillation and a slow decay of the number of atoms in the excitation modes, $N_B$. The first minimum of $N_B$ is at $T_h\approx3$\,ms, the maximum at 6\,ms, and exponential decay times are approximately $15$\,ms. We find that $T_h$ depends on lattice depth and interaction strength, but not on the initial occupation strength of the excitation mode. Measured values of $T_h$ show a similar trend for varying lattice depths as $2\pi/\omega_q$ (red circles and blue squares in Fig.\,\ref{fig:StaticMIHealing}(c)).

We speculate that the time evolution of $N_B$ in Fig.\,\ref{fig:StaticMIHealing} is caused by an oscillation between two configurations of a density wave. After the Bragg pulse, the initial state consists of a carrier wave and two excitation modes of similar strength and quasimomentum $\pm q$. Interfering these modes creates a density modulation of the carrier wave with wavelength $\lambda_q = 2\pi/q$ (sketch in Fig.\,\ref{fig:StaticMIHealing}(c)). This density wave grows due to MIs for $|k|>k_L/2$ (Sec.\,\ref{sec:Exp_GrowthMI}), but it decays for a stable system with $|k|<k_L/2$. For those stable carrier momenta, repulsive interactions cause the wave to spread and to regenerate at a shifted position $\lambda_q/2$. As a result, the time evolution of $N_A$ shows an oscillation with period $2\pi/\omega_q$ between these two configurations of the density wave.

The damping of the oscillations can be attributed to our trapping potential and the finite system size. In our experimental setup, the trapping potential breaks the translational symmetry and lifts the energy degeneracy for the two configurations of the density wave, which results in the dephasing and decay of the oscillations. However, the measured timescale of 10 to 20\,ms for damping is shorter than expected for our trap periods of approximately 100\,ms, and we did not observe dephasing or heating in the momentum profiles in Fig.\,\ref{fig:StaticMIHealing}(a)).

%***********************
% Instabilities with driving
%***********************
\subsection{\label{sec:DrivenInstabilities} MIs and PIs with periodic driving}

We separately examine in this section both types of instabilities at intermediate driving frequencies. For MIs, we demonstrate the impact of micromotion on excitation modes by observing a complex pattern of growth and decay as the carrier wave cycles through the Brillouin zone (Sec.\,\ref{sec:Exp_MIDriven})). For PIs, we provide evidence of their existence in a parameter regime where MIs are not possible (Sec.\,\ref{sec:Exp_PIDriven}).

\subsubsection{\label{sec:Exp_MIDriven} Modulational instabilities.}

In driven systems, MIs occur when the carrier wave periodically passes through regions of the Brillouin zone with positive and negative effective mass. As we demonstrated in Sec.\,\ref{sec:Exp_GrowthMI} and \ref{sec:Exp_GrowthStabMI}, MIs grow in unstable regions but oscillate and decay in stable regions of the Brillouin zone. In this section, we show that, for periodic driving, both effects combine to create a complex oscillation of the excitation modes.

\begin{figure}[t]
\centering
  \includegraphics[width=0.49\textwidth]{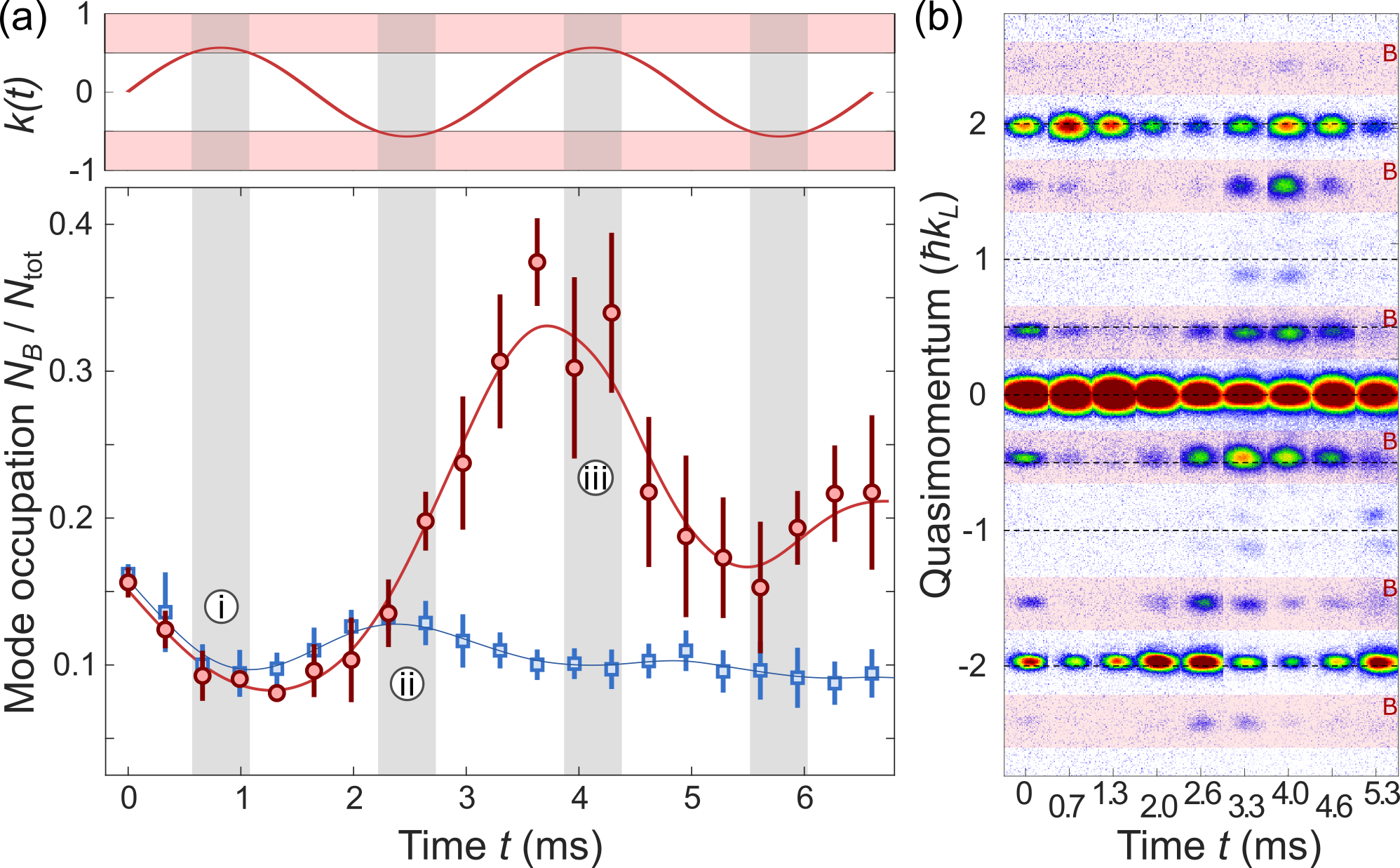}  % fig10

  \caption{Excitations growth due to MIs with driving. (a) Time evolution of a seeded excitation mode without driving (blue squares) and with driving (red circles), $q=0.48\,k_L, V=6.3\,E_r$, $a=105\,a_0$, $T_D=3.3\,ms$. Top panel: Micromotion for $K=1.8$, $k_0=0$. Critical regions of the BZ are indicated by red patches and time intervals of $k(t)$ in those critical regions by gray patches. The lines next to data sets are smoothing splines to guide the eye. (b) Averaged absorption images for the driven system. We observe no significant heating during the initial evolution time. \label{fig:DrivenMIGrowth}} % data 31.07.22, $q=0.48\,k_L, V=6.3\,E_r$, $a=105\,a_0$, $\omega_z =2\pi\times 21$\,Hz
\end{figure}

In a first step, we applied a Bragg pulse that generated a pair of excitation modes with 15\% of the carrier wave. Without driving, the resulting density wave was unstable for $k=0$ and exhibited damped oscillations of the mode occupation with a first minimum at $T_h\approx1.1$\,ms (blue squares in Fig.\,\ref{fig:DrivenMIGrowth}(a)). In a second step, we added a driving force with period $T_D$ and observed its impact on the mode occupation (red circles in Fig.\,\ref{fig:DrivenMIGrowth}(a)). For a driving strength $K=1.8$ and $T_D\approx3T_h$, the micromotion crossed briefly into regions with negative effective mass whenever the non-driven oscillation of the excitation mode was close to a maximum or minimum. Red and gray patches in Fig.\,\ref{fig:DrivenMIGrowth}(a) indicate intervals in quasimomentum space and time with negative effective mass.

The first crossing of the micromotion into an unstable region of the Brillouin zone (marked by (i) in Fig.\,\ref{fig:DrivenMIGrowth}(a)) has little effect on the excitation mode, as the mode's occupation is small and decreasing. The driving force even reduces the mode occupation between 1\,ms and 2\,ms compared to the non-driven system. However, when the occupation of the non-driven mode is at a maximum at the second crossing (ii), the driven excitation mode grows strongly by almost a factor three. At the third crossing (iii), as the mode occupation begins to decrease, the oscillation amplitude is reduced again. We found that the mode's oscillation amplitude continued to fluctuate with a complex interplay between micromotion and mode occupation.

We did not observe a significant increase of the number of thermal atoms during the first 5\,ms (Fig.\,\ref{fig:DrivenMIGrowth}(b)) which suggests that the process was mostly coherent. For MIs, excitation modes must occur in pairs with opposite momenta $(+\hbar q,-\hbar q)$ to conserve the total momentum. Also the energy is conserved due to the symmetry of Eq.\,(\ref{eq:ExTrombettoni}) for which the real parts of $\omega_{+q}$ and $\omega_{-q}$ have opposite signs (interval with overlapping blue lines in Fig.\,\ref{fig:StaticEnergy}(b)). As a result, pairs of excitation modes can grow for $|k|>0.5k_L$ without changing the system's energy (e.g. Fig.\,\ref{fig:StaticMIGrowth}(b)). However, we expect that the observed oscillations result in a slow heating of the system as the micromotion drags existing modes through the Brillouin zone. Further studies are needed to evaluate this mechanism on longer time scales. The mechanism is different for PIs that can be created at $|k|<0.5k_L$ when the total energy of the modes is positive.

\subsubsection{\label{sec:Exp_PIDriven} Parametric instabilities}

\begin{figure}[t]
\centering
  \includegraphics[width=0.49\textwidth]{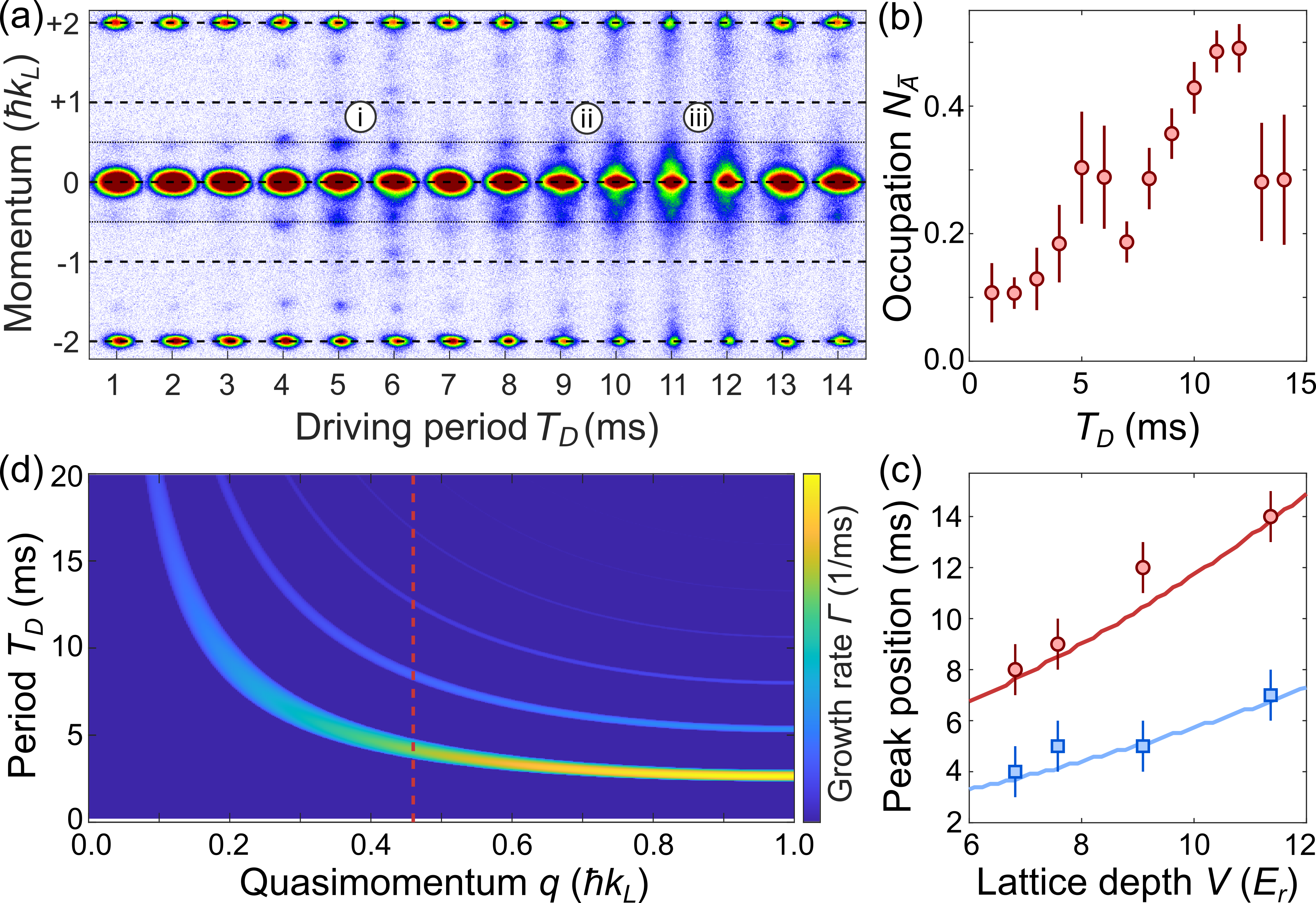} % fig11
  \caption{Measurement of parametric instabilities. (a) Absorption images with momentum distribution after driving for approximately 30\,ms ($a=73\,a_0$, $K=1.3$, $V = 9\,E_r$, $N_\text{tot}=3.4\cdot10^4$). An initial Bragg pulse seeds excitation modes at $q=0.46\hbar k_L$. (b) Number of atoms that are not in the ground state, $N_{\bar{A}}$, for the data in (a). (c)  Measured driving periods of the first (blue squares) and second (red circles) peak of $N_{\bar{A}}$ for varying lattice depths. Solid lines show calculated peak positions. (d) Numerically calculated growth rates using Eq.\,(\ref{eq:BogoliubovGennes}) and parameters in (a). The dashed line indicates the quasimomentum of the exited excitation mode. \label{fig:Parametric}} % data 24.11.22,  7.2A, dimple 0.8mW, 30mW, K drive 2.2 kHz, 70ms expansion V = 9E
\end{figure}

Weak driving strengths seem to be ideally suited for the search of PIs. For $K<\pi/2$, the micromotion does not cross into regions of the Brillouin zone with negative effective mass and MIs are not possible. However, we did not detect the growth of excitation modes in this region in our measurements for the stability diagram (see Sec.\,\ref{sec:InstabilityDiagram}). Instead, we employed a different approach by demonstrating the reduced decay of existing modes due to PIs. Using a Bragg pulse, we created excitations at $q=0.46 k_L$ and monitored their decay.

For most driving periods, those excitation modes were no longer visible in the momentum profiles after 30\,ms (Fig.\,\ref{fig:Parametric}(a)). However, some weak excitations persisted for $T_D\approx5\,$ms (i) and $T_D\approx10$\,ms (ii). We determined the number of atoms in the ground state, $N_A$, from Gaussian fits to the peaks at $0,\pm 2\hbar k_L$ , and calculated the relative number of excited atoms as $N_{\bar{A}} = 1-N_A/N_\text{tot}$. The two peaks of $N_{\bar{A}}$ in Fig.\,\ref{fig:Parametric}(b) match well to the observed excitation modes in the absorption images. Figure\,\ref{fig:Parametric}(d) compares the measured resonance positions with the calculated growth rates. The initially excited mode at $q=0.46 k_L$ is indicated by the dashed red line. We measured the two resonance positions for varying lattice depth and compared them to the calculated growth rates (Fig.\,\ref{fig:Parametric}(c)). We observed good agreement between the measurements and the predictions, except for the large strength of the second resonance at $T_D\approx 10\,$ms. The width and height of its peak in Fig.\,\ref{fig:Parametric}(b)) are likely due to the coupling with other excitation modes at smaller $q$, which can also be seen in the absorption images at (iii).

This measurement demonstrates the existence of PIs, but their growth rates seem too small to excite the system within 30\,ms by random seeding. We find for the parameters in Fig.\,\ref{fig:Parametric}(a) that the maximum calculated growth rate for weak driving (only PIs) is a factor of 3 smaller than the growth rate for stronger driving with $\pi/2<K<2.4$ (PIs and MIs). This variation in growth rates may be due to the different coupling mechanisms between the carrier and excitation modes for PIs and MIs. PIs require an oscillating driving force to excite the carrier wave and add energy, whereas MIs are achieved by periodically switching the instability on and off without adding energy.

\section{\label{sec:InstabilityDiagram} Interpretation of the stability diagram}

\begin{figure}[t]
\centering
  \includegraphics[width=0.49\textwidth]{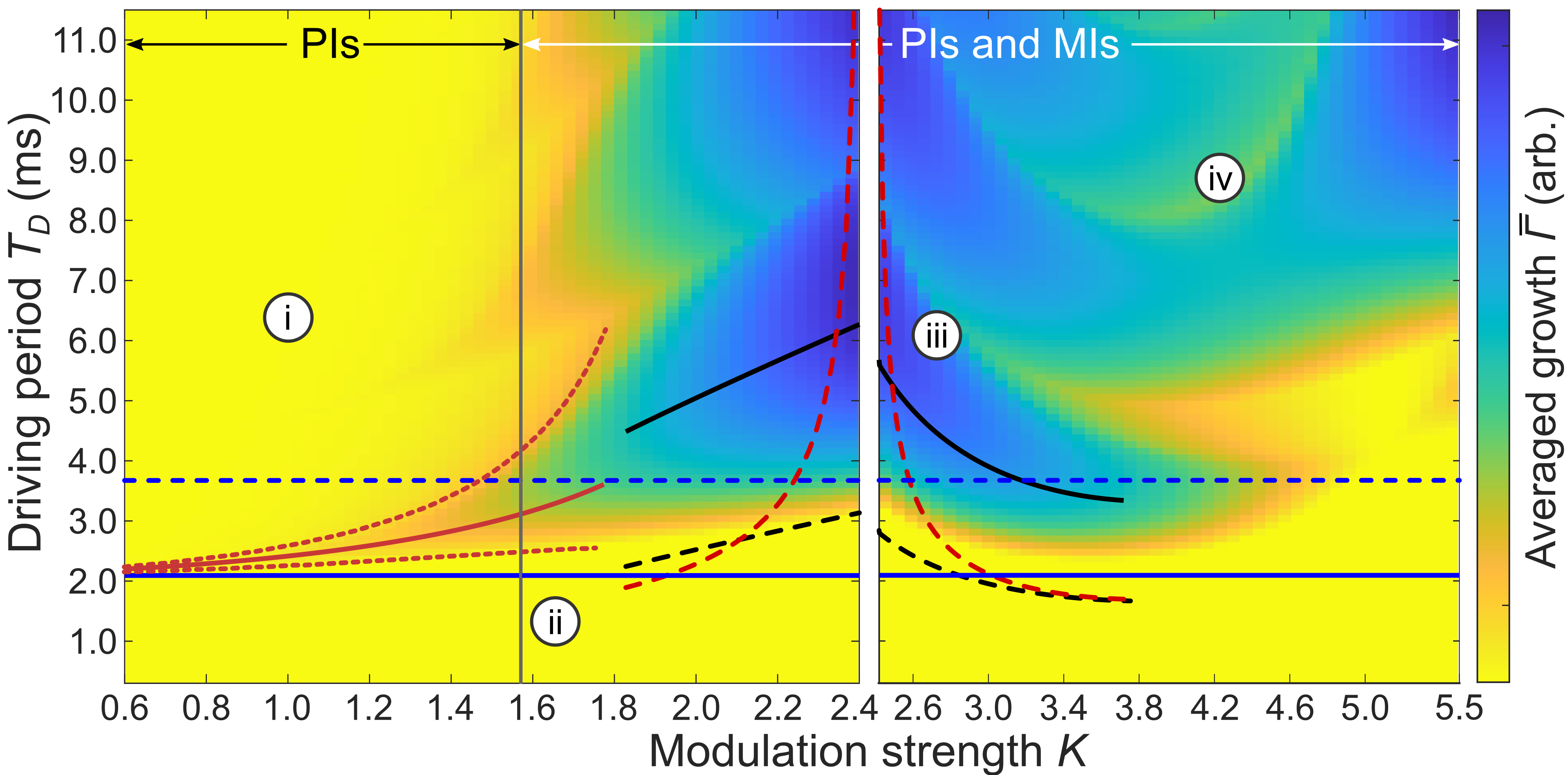}  % fig12
  \caption{ Stability diagram. Numerically calculated stability using $\overline{\Gamma}(K,T_D)$ ($V=8.8\,E_r$, $U/J=22$). Solid red and black lines show $2\pi/\omega^f_q$ and $2\pi/\omega^s_q$ $q=\hbar k_L$, dashed lines show half of the corresponding values. The black vertical line at $K=\pi/2$ separates weak from strong driving regimes. Solid and dashed horizontal blue lines provides $2\pi/\omega_\text{max}$ and $h/8J$, respectively. Regions indicated by Roman numerals are discussed in the text.
  \label{fig:OverviewTheory}}
\end{figure}

Finally, we bring together our findings from earlier sections to interpret the measurement of the stability diagram in Fig.\,\ref{fig:StabilityIntro}. We compare our measured results to numerically calculated growth rates $\Gamma_q(K,T_D)$ which are determined using Eq.\,(\ref{eq:BogoliubovGennes}) \cite{creffield2009} (Fig.\,\ref{fig:OverviewTheory}). The randomness of the excitations' quasimomentum (see Sec.\,\ref{sec:Exp_MomentumMI}) and the fast coupling between excitation modes make it challenging to determine $q$ for the initial excitation. Instead of using the most unstable mode, we account for this spread of quasimomenta by averaging the growth rate over the Brillouin zone, $\overline{\Gamma} = \braket{\Gamma_q(K,T_D)}_q$. The parameter $\overline{\Gamma}$ is no longer a growth rate but provides an indicator for the system's stability. Yellow colors in Fig.\,\ref{fig:OverviewTheory} correspond to a stable system with low values of $\overline{\Gamma}$, while blue colors indicate large $\overline{\Gamma}$ and low stability. The structure of $\overline{\Gamma}$ along the $T_D$ axis results from higher-order resonances (see Fig.\,\ref{fig:IllustratePIs}).

Roman numerals in Fig.\,\ref{fig:OverviewTheory} indicate the same characteristic regions as in Fig.\,\ref{fig:StabilityIntro}. We did not observe any excitation modes in the weak driving regime (i) as discussed in Sec.\,\ref{sec:Exp_PIDriven}, although numerical calculations predict the creation of PIs. We speculate that the PIs' growth rates are insufficient for our observation times and for random seeding. The system starts to become unstable when we cross into the region of strong driving strength at $K=\pi/2$ (vertical gray line in Fig.\,\ref{fig:OverviewTheory}). In addition, the measurement shows that the system is stable for small values of $T_D$ at all driving strengths (region (ii)), which corresponds to the fast driving regime. The energy threshold $\hbar \omega_{max}$ (Sec.\,\ref{sec:InstabilitiesTheory}) predicts the fast driving limit correctly (solid blue line), however, the condition is strict and not useful to describe the transition line between stable and unstable driving periods. As a reference, we also provide the time scale $h/8J$ of twice the width of the lattice band (blue dashed lines).

The calculated parameter $\overline{\Gamma}$ provides a good qualitative estimate for the observed regions of stability. The system is least stable for $K\approx 2.5$ in region (iii), and the transition line increases to larger values of $T_D$ at point (iv). Even the small cusp in the transition line at $K\approx 2.4$ matches between prediction and experiment. However, the agreement is only qualitative and values of $T_D$ for the transition line are smaller in the calculation than in our measurement. For example, the transition from stable to unstable occurs at $K=2.4$ for $T_D\approx 6$\,ms in the experiment and for 3\,ms in the calculation. In general, the system is more stable in our experimental measurement than predicted by the calculation, which might be due to the short observation time. Using longer driving durations would improve our measurement sensitivity, but requires the consideration of additional time scales, such as trapping periods.

We provide several analytical solutions from \cite{bukov2015,lellouch2017} for comparison (lines in Fig.\,\ref{fig:OverviewTheory}). For weak driving strengths, the resonance condition for PIs is well approximated by $\omega_D = \omega^f_{q=k_L}$ (solid red line) with a boundary $\omega_D = \omega^f_{k_L} \pm 4 J J_2(K)U/(\hbar^2\omega^f_{k_L})$, where $J_2$ is the second-order Bessel function (dotted lines) \cite{lellouch2017}. For strong driving strength and $K\approx2.4$, $\omega_D=\omega^s_{k_L}$ provides a good approximation for the maximum growth rate (solid black line) and $\omega_D=2\omega^s_{k_L}$ indicates the boundary between the unstable zone and the stable fast driving regime (dashed black line). The boundary initially suggested in \cite{bukov2015}, $\omega_D = 2\omega^f_{k_L}$, is indicated by a dashed red line.

\begin{figure}[t]
\centering
  \includegraphics[width=0.49\textwidth]{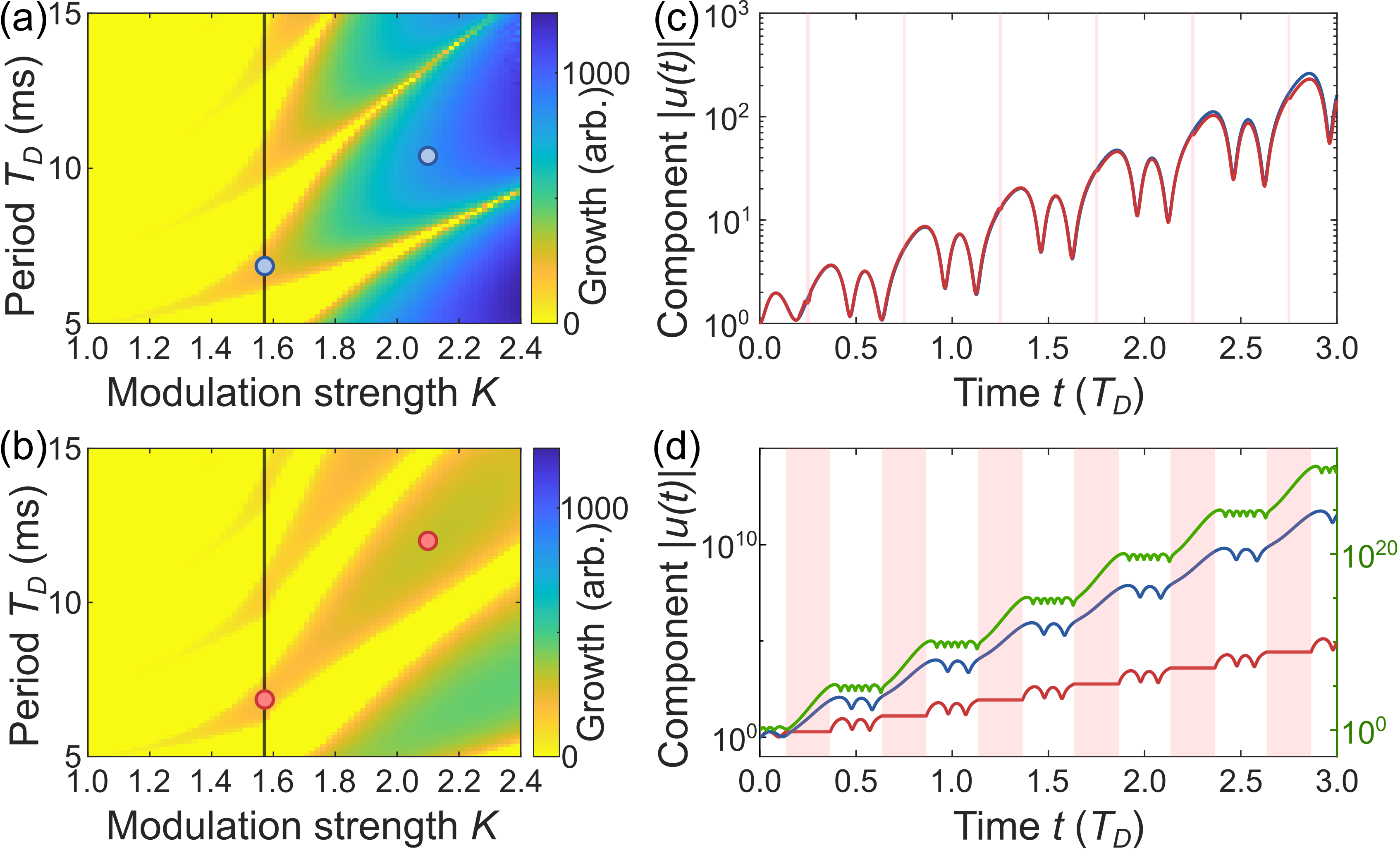}  % fig13
  \caption{Comparison of MIs and PIs. (a) Calculated growth rate for mode $q=k_L$ with the same parameters as in Fig.\,\ref{fig:OverviewTheory}. (b) Calculated growth rate with interactions switched-off whenever the micromotion passes through a region with negative effective mass. Circles indicate values of $(K, T_D)$ for lines with matching colors in panels (c,d). (c) Time evolution $|u(t)|$ for an initial state $(u=1,v=0)$ and $K=1.572$ with switched interactions (red line) and without (blue line). (d) Time evolution for $K=2.1$ with (red line) and without (blue line) switched interactions, and in the slow driving regime with $T_D=30.6\,$ms (green line).
  \label{fig:CompareGrowth}}
\end{figure}

Distinguishing between MIs and PIs in Fig.\,\ref{fig:OverviewTheory} is challenging for intermediate driving frequencies because both types of instabilities can occur simultaneously. To identify the contributions of both mechanisms, we calculated the growth rate $\Gamma_{q=k_L}(K,T_D)$ as before using Eq.\,(\ref{eq:BogoliubovGennes}). However, we switched off interactions with $U=0$ whenever the micromotion passes through a region with negative effective mass. This approach allowed us to compare the growth caused by both instabilities (Fig.\,\ref{fig:CompareGrowth}(a)) to the growth caused only by PIs (Fig.\,\ref{fig:CompareGrowth}(b)). While the separation of MIs and PIs is not perfect, as the method reduces the energy of excitation modes and induces a shift of resonance frequencies towards larger values of $T_D$ in Fig.\,\ref{fig:CompareGrowth}(b), it still provides a good indication of the significant increase of $\Gamma$ due to MIs.

We compare the time evolution of $|u(t)|$ for an initial state $(u=1,v=0)$ with and without switched interactions (red and blue lines in Fig.\,\ref{fig:CompareGrowth}(c,d)). We find that PIs dominate the time evolution when crossing just into the strong driving regime ($K=1.572$, Fig.\,\ref{fig:CompareGrowth}(c)) and the growth is mostly independent of the negative effective mass regions (red patches). For a stronger driving strength ($K=2.1$, Fig.\,\ref{fig:CompareGrowth}(d)), PIs still contribute to the growth (red line), but most of the growth is caused by MIs (blue line). As we approach the slow driving limit (green line, $T_D=30.6\,$ms), the influence of PIs can be neglected, and the system is stable except for those time intervals when the micromotion passes through regions with a negative effective mass.

%***********************
% Conclusion
%***********************

\section{\label{sec:conclusion} Conclusions}

In conclusion, we studied the stability of driven matter waves with repulsive interactions in a 1D lattice. By measuring the momentum distribution of the matter waves after a fixed hold time, we identified stable and unstable regions based on the driving period $T_D$ and strength $K$. Our results emphasize the importance of making a clear distinction between modulation and parametric instabilities in driven systems. Parametric instabilities arise when the drive resonantly couples to excitation modes, while modulational instabilities stem from inherent properties of the medium, such as attractive interactions or a negative effective mass. Our key findings are summarized as follows.

Modulational instabilities are well suited for describing intra-band excitations in the limits of fast and slow driving frequencies. We demonstrated for the fast driving limit that phonon energies and stability criteria map directly to the non-driven system. This enables the use of established concepts for the stability analysis of interacting many-body states in Floquet-driven systems. The description is applicable for the initial growth of excitation modes while perturbation theories remain valid.

Of particular interest is the regime of intermediate driving frequencies, where both modulational and parametric instabilities are present. The micromotion of the driven wave packet periodically crosses into regions of the Brillouin zone that are unstable due to modulational instabilities, resulting in a complex growth and decay of excitation modes. Directly demonstrating parametric instabilities was challenging due to their low growth rates. As an alternative, we demonstrated a reduced decay of existing excitation modes at the driving parameters where only parametric instabilities are expected.

We compared the measured stability diagram with predictions for the growth of excitation modes based on the Bogoliubov-de Gennes equations. There is a good qualitative agreement of stable and unstable zones, however, our experimental system was more stable than predicted and the stable fast-driving regime extended to larger driving periods. We found that the time-averaged energies $\hbar \omega^f_q$ and $\hbar \omega^s_q$ provide good estimates for the position of resonances and boundaries at weak driving strength and for $K\approx 2.4$.

Our results have important implications for experiments using Floquet-engineered potentials, particularly for applications that require minimal excitations and heating. By identifying stable and unstable parameter regions, we can predict which experimental conditions will lead to unwanted excitations. Energy transfer and excitation mechanisms are different for modulational and parametric instabilities, and a clear distinction between these two types of instabilities will be instrumental to study the system over longer time scales as required, e.g., for the creation of discrete time crystals \cite{sacha2017} or dynamical gauge fields \cite{zohar2015}.

\vspace{5ex}

We acknowledge support by the EPSRC through a New Investigator Grant (EP/T027789/1), the Programme Grant DesOEQ (EP/P009565/1), and the Quantum Technology Hub in Quantum Computing and Simulation (EP/T001062/1). CEC was supported by the Universidad Complutense de Madrid through Grant No. FEI-EU-19-12.

\renewcommand{\thefigure}{S\arabic{figure}}
\setcounter{figure}{0}
\refstepcounter{figure}\label{fig:StabilityExamples}
\refstepcounter{figure}\label{fig:StabilityImages}

%\clearpage
%\newpage

\bibliography{./Modulational_Instability, ./OurPapers}

\clearpage
\newpage

%***********************
% supplemental materials
%***********************

\renewcommand{\thefigure}{S\arabic{figure}}
\setcounter{figure}{0}

\section*{Supplemental Material \label{sec:supplemental}}

\begin{figure}[t]
\centering
  \includegraphics[width=0.49\textwidth]{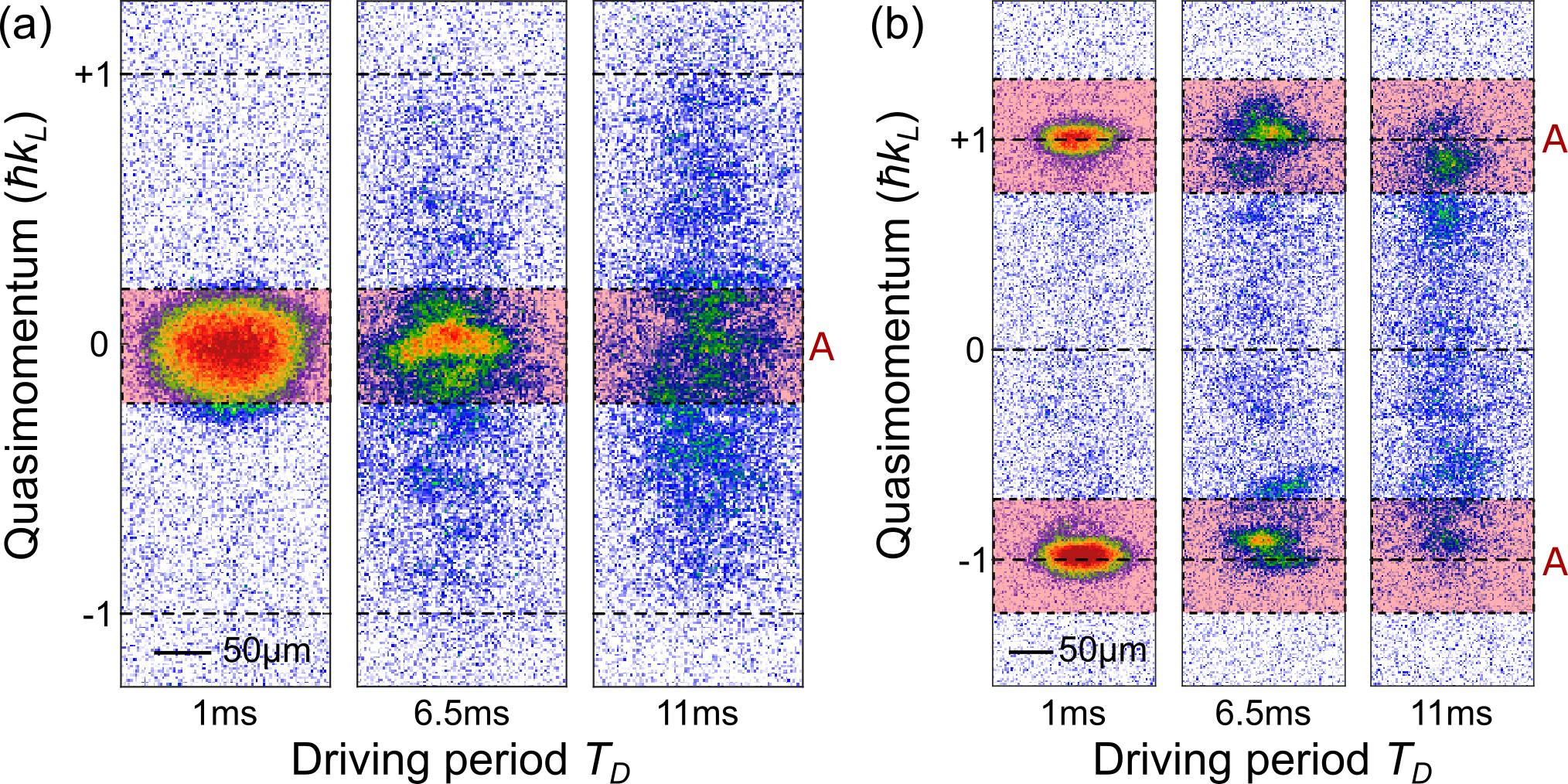}
  \vspace{-2ex}
  \caption{Examples of absorption images of excitation modes. Images show the quasimomentum distribution after driving for approximately 30\,ms with driving strength (a) $K = 2.31$ and (b) $K = 4.15$. The number of atoms in excitation modes, $N_A/N_\text{tot}$, is calculated as the ratio of atoms in regions A (red patches) and the total atom number. \label{fig:StabilityExamples}}
\end{figure}

\begin{figure}[t]
\centering
  \includegraphics[width=0.49\textwidth]{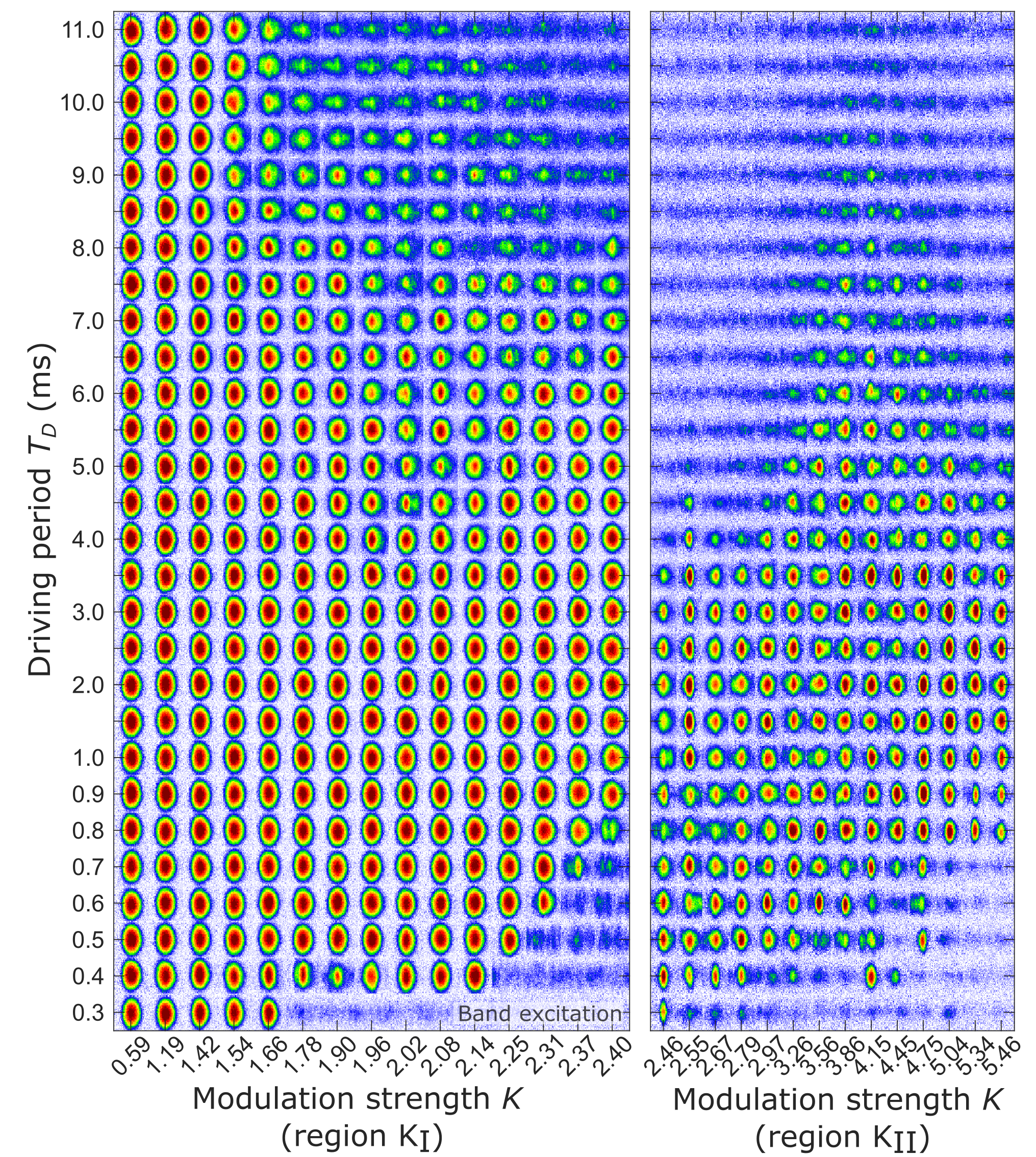}
  \vspace{-2ex}
  \caption{Absorption images for the stability diagram in Fig.\,\ref{fig:StabilityIntro}. The images are averaged over at least 4 repetitions. For a direct comparison of images in intervals $K_I=0-2.4$ and  $K_{II}=2.4-5.5$, we combine the two density peaks of the ground state in interval $K_{II}$ by summing upper and lower half of the images. \label{fig:StabilityImages}}
\end{figure}

This section provides additional parameters and example images for the stability measurement in Fig.\,\ref{fig:StabilityIntro}. A BEC is prepared with approximately 40,000 atoms at a scattering length $a=104\,a_0$ in crossed beam dipole trap with frequencies $\omega_{x,y,z} = 2\pi\times (18,21,10)$\,Hz. We increase the vertical 1D lattice potential $L_1$ in 150\,ms to load the atoms to a lattice depth $V=8.8\,E_r$ and modulate the lattice velocity for a duration $t$. The quasimomentum distribution of the atoms is measured by absorption imaging after 75\,ms of levitated expansion and 2\,ms of free time-of-flight.

The wave packets are initially prepared in ground states, which are centred at $k_0=0$ in interval $K_I=0-2.4$ and at $k=\pm k_L$ in interval $K_{II}=2.4-5.5$. For $k_0=\pm k_L$, we accelerate the atoms to the edge of the Brillouin zone by reducing the levitating magnetic field gradient over 6.2ms. Figure \ref{fig:StabilityExamples} provides single absorption images as examples for the evolution of the wave packet for different driving periods $T_D$. The wave packets are stable for fast driving with $T_D=1\,$ms, but fragment and spread for larger driving periods. We quantify the stability of the system by measuring the relative atom number close to the momentum peaks of the ground state, $N_A(\Delta t)/N_\text{tot}(\Delta t)$ (red patches in Fig.\,\ref{fig:StabilityExamples}).

The precise driving duration is adjusted to that multiple of the period $T_D$ which is closest to 30\,ms. We confirmed that the residual time variations do not result in qualitative changes when $T_D$ is varied from 0.3 to 11\,ms. Using fixed hold times instead of a fixed number of oscillations is more applicable to quantum simulation experiments. The hold time of 30\,ms was chosen to be close to a quarter of the longitudinal trap period because our previous research showed that the trapping potential has a significant impact on the evolution of driven systems \cite{dicarli2019b}. Keeping the hold time short reduces the effects of longitudinal and radial trapping potentials that have frequencies close to 10\,Hz and 20\,Hz, respectively.

Figure \ref{fig:StabilityImages} provides an overview of the absorption images in intervals $K_I$ and $K_{II}$ for $T_D=0.4-11$\,ms. Each image is averaged over at least 4 realizations. The density profiles of the ground states show one peak in interval $K_I$ and two peaks in $K_{II}$. For a better comparison of the intervals in Fig.\,\ref{fig:StabilityImages}, we add the density distributions at the edges of the Brillouin zone for $K_{II}$. The lower half and upper half of the images are added together. The images already provide a good indication of unstable regions. Instabilities with fast driving frequencies $T\lesssim0.6$\,ms can be associated with the coupling to higher bands which has been studied in \cite{song2022}. Here, we focus on instabilities for slower driving (blue regions at top of figure), which can be caused by modulational instabilities or parametric instabilities.

\vfill

\end{document}